\documentclass[
aps,
prb,
floatfix,
superscriptaddress,
twocolumn,
reprint, 
longbibliography
]{revtex4-2}

\usepackage{xcolor}
\usepackage[colorlinks=true,allcolors=blue]{hyperref}

\usepackage{amsmath}
\usepackage{amsfonts}
\usepackage{graphicx}
\usepackage{bm}
\usepackage{dsfont}
\usepackage{xfrac}
\usepackage{siunitx}
\usepackage{nicefrac} 
\usepackage{hyperref}
\usepackage[normalem]{ulem}
\usepackage{braket}
\usepackage{float}
\usepackage{multirow}

\usepackage{tikz}
\usepackage{verbatim}
\usepackage{bookmark}
\usetikzlibrary{decorations.pathmorphing}
\usetikzlibrary{arrows}
\usepackage{nicefrac}

\graphicspath{./}

\begin{document} 

\title{Renormalized density matrix downfolding: a rigorous framework in learning emergent models from \textit{ab initio} many-body calculations}

\author{Yueqing Chang}
\email{yueqing.chang@rutgers.edu}
  \affiliation{Center for Materials Theory, Department of Physics \& Astronomy, 
Rutgers University, Piscataway, New Jersey 08854, USA}
\author{Sonali Joshi}
\affiliation{Department of Physics, University of Illinois at Urbana-Champaign, Urbana, Illinois 61801, USA }
\author{Lucas K. Wagner}
\email{lkwagner@illinois.edu}
\affiliation{Department of Physics, University of Illinois at Urbana-Champaign, Urbana, Illinois 61801, USA }

\begin{abstract}
    We present a generalized framework, renormalized density matrix downfolding (R-DMD), to derive systematically improvable, highly accurate, and nonperturbative effective models from \textit{ab initio} calculations. 
    This framework moves beyond the common role of \textit{ab initio} calculations as calculating the parameters of a proposed Hamiltonian.
    Instead, R-DMD provides the capability to decide whether a given effective Hilbert space can be identified from the \textit{ab initio} data and assess the relative quality of \textit{ansatz} Hamiltonians.
    Any method of \textit{ab initio} solution can be used as a data source, and as the \textit{ab initio} solutions improve, the resultant model also improves. 
    We demonstrate the framework in an application to the downfolding of a hydrogen chain to a spin model, in which we find the interatomic separations for which a nonperturbative mapping can be made even in the strong coupling regime where standards methods fail, and compute a renormalized spin model Hamiltonian that quantitatively reproduces the \textit{ab initio} dynamics.
\end{abstract}

\maketitle

\section{Introduction}

A fundamental objective of condensed matter physics is to classify the diverse states of matter that emerge from the common makeup of electrons and nuclei. 
This endeavor has been principally guided by the renormalization group approach~\cite{glazek_renormalization_1993, glazek_perturbative_1994, RevModPhys.47.773}, which provides a systematic framework for the problem of quantum impurity with a few effective degrees of freedom that couple to a bath. 
The crucial tool is the effective Hamiltonians or Lagrangians, which are often written in terms of quasiparticles.
Despite the utility, the applications of non-perturbative renormalization group techniques remain challenging, particularly in the study of complex materials like cuprates~\cite{white_density_1992, maier_d_2000, becca_stripes_2001, sorella_superconductivity_2002, anisimov_first_2002, spanu_magnetism_2008, ashkenazi_theory_2011, hirayama_derivation_2013, qin_absence_2020, jiang_ground-state_2021, jiang_pairing_2022}, pyrochlore iridates~\cite{witczak-krempa_correlated_2014, rau_spin-orbit_2016}, spin liquid candidates~\cite{banerjee_neutron_2017, savary_quantum_2017, zhou_quantum_2017}, and twisted bilayer graphene~\cite{andrei_graphene_2020, andrei_marvels_2021}.
The intricacy of these materials often prevents effective Hamiltonians from being fully determined through experimental means alone.

The state of the art in \emph{downfolding} from \textit{ab initio} calculations is to construct an effective Hamiltonian from density functional theory (DFT), with interactions added either through a constrained random phase approximation (cRPA)~\cite{aryasetiawan_frequency-dependent_2004,aryasetiawan_calculations_2006} and beyond~\cite{scott_rigorous_2024}, or $GW$ approximation~\cite{hedin_new_1965, aryasetiawan_gw_1998, hirayama_derivation_2013, romanova_dynamical_2023, canestraight_efficient_2024}.
However, since they are based on DFT, these downfolding techniques are limited by their mean-field ingredients~\cite{cohen_insights_2008, williams_direct_2020} and require careful considerations, from the choice of the basis to the double counting correction scheme~\cite{chang_downfolding_2024}; therefore, their accuracy is less reliable for strongly correlated systems. 
For example, the correct interacting models are crucial for determining the topological and magnetic properties of several pyrochlore iridates~\cite{zhang_metal_2017, wang_electron_2017, wang_weyl_2017, liu_magnetic_2021} but yet to be systematically derived.
Another example is the Fe defects in AlN, where DFT+cRPA with various double counting corrections fail to reproduce the correct ordering of the lowest few excited states, and therefore demands more accurate \textit{ab initio} many-body data~\cite{muechler_quantum_2022}.

Recent advancements in many-body \textit{ab initio} methods have demonstrated the potential to achieve high-accuracy solutions even in strongly correlated systems. 
Techniques such as multi-reference quantum chemistry methods~\cite{huron_iterative_1973, ruedenberg_are_1982, cimiraglia_recent_1987, holmes_heat-bath_2016}, various flavors of quantum Monte Carlo~\cite{baroni_reptation_1999, foulkes_quantum_2001, zhang_quantum_2003, booth_fermion_2009, foyevtsova_abinitio_2014}, and density matrix renormalization groups~\cite{white_density_1992, white_ab_1999, chan_highly_2002} have adapted heuristics to achieve systematically improvable approximations of the ground states.
These methods have been shown to achieve converged results, albeit constrained to limited system sizes of up to 30 electrons~\cite{shepherd_convergence_2012, williams_direct_2020, motta_ground-state_2020, benali_toward_2020}.
A practical way to explore larger systems and extended length scales is via the lower energy effective theories. 
However, to date, a systematic approach to bridging these high-accuracy many-body methods with lower energy-effective theories remains elusive.

In this work, we introduce a rigorous and non-perturbative method to compute downfolded models using accurate many-body \textit{ab initio} wave functions, called renormalized density matrix downfolding (R-DMD). 
The quality of the model is measured by its ability to reproduce the \textit{ab initio} data, allowing for systematic improving models based on obtaining the best fit for the \textit{ab initio} data.
Therefore, the method is amenable to machine learning techniques, especially feature selection techniques.
Renormalization of physical quantities, such as effective spin, is obtained as a result of the downfolding and can be dynamical, in contrast to standard downfolding techniques~\cite{szilva_quantitative_2023}. 
The inherent non-perturbative treatment of interactions in many-body \textit{ab initio} calculations endows R-DMD with the capability to accurately describe the correlated regime away from the atomic limit, where nearest-neighbor hopping is comparable to the on-site Hubbard repulsion ($t\lesssim U$).

We demonstrate R-DMD using the strongly correlated hydrogen chain, in which we identify the regions of phase space where the \textit{ab initio} system can be mapped onto a spin model and construct a complete non-perturbative renormalization to a spin model where possible.
The spectral residual obtained using R-DMD is minimal compared to other downfolding methods, including symmetry-broken DFT, Schrieffer-Wolff transformation, spectral-fitting DMD, and unrenormalized DMD.
The accuracy of symmetry-broken DFT is limited by its input \textit{ab initio} data.
In the atomic limit, all methods perform well, but in the correlated regime away from the atomic limit, Schrieffer-Wolff transformation fails due to its perturbative nature, and unrenormalized DMD performs poorly because the spin operator is heavily renormalized.
We highlight that R-DMD works well in both cases~\footnote{For the simple case of \textit{ab initio} hydrogen chain, spectral-fitting DMD also performs well in both cases, but could scramble the physics for more complicated system}.

\begin{figure}
	\includegraphics[width = 0.45\textwidth]{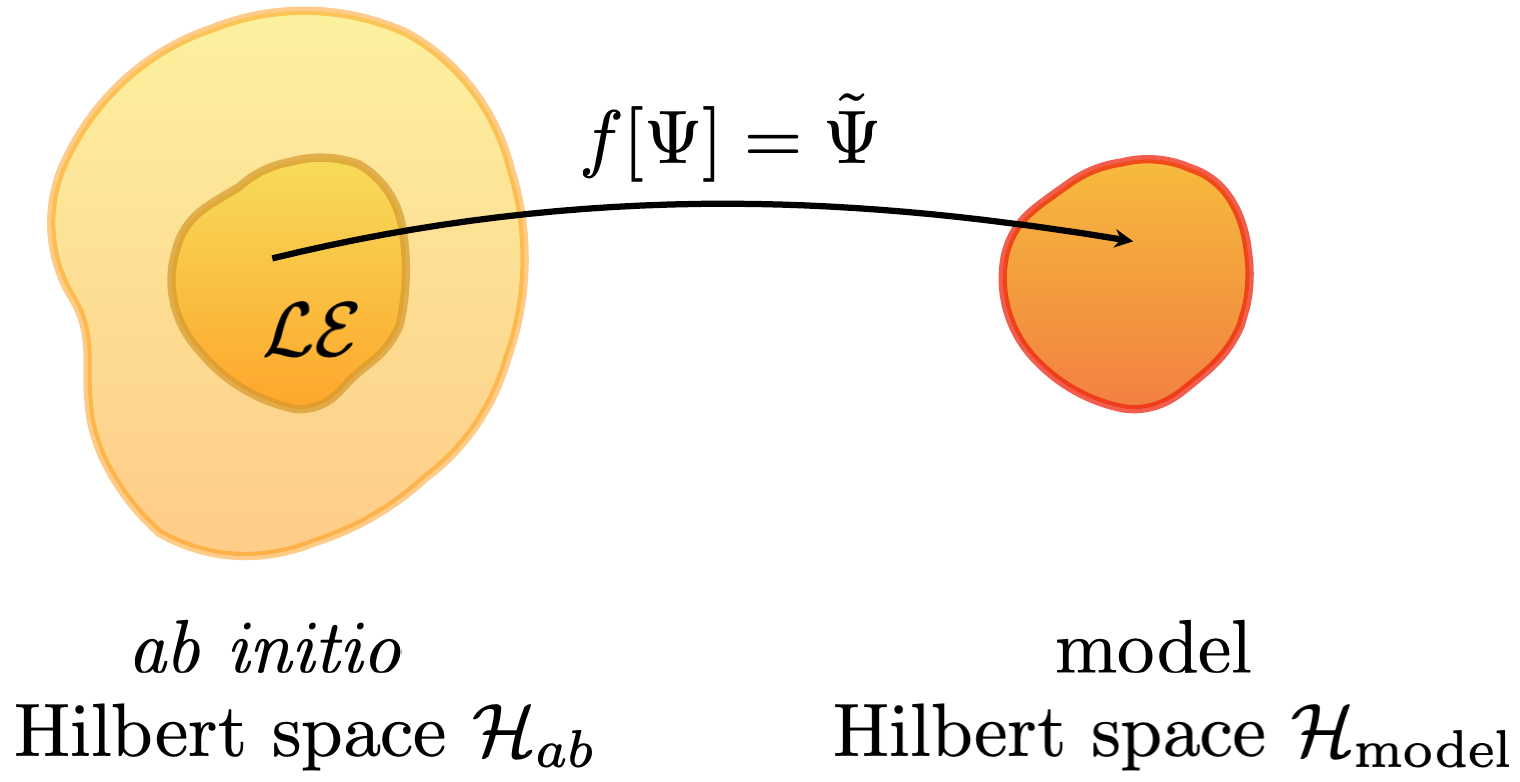}
	\caption{A 1-1 mapping between the \textit{ab initio} emergent low-energy Hilbert space and the model Hilbert space enables \textit{ab initio} dynamics to be reproduced by an effective model.}
	\label{fig:one-to-one_mapping}
\end{figure}

To help understand the choices in downfolding, in Fig~\ref{fig:one-to-one_mapping}, we outline the general procedure. 
A subspace of the \textit{ab initio} Hilbert space $\mathcal{H}_{ab}$, called $\mathcal{LE}$ is chosen to represent the low-energy space.
A 1-1 mapping $f$ is constructed between $\mathcal{LE}$ and some model Hilbert space $\mathcal{H}_{\text{model}}$.
We start with the following general considerations, which are easy to show: 
\begin{enumerate}
    \item[(1)] If $\mathcal{LE}$ is spanned by a set of energy eigenstates, then dynamics within that space are closed (i.e., an effective Hamiltonian is sufficient). If it is not, then the resulting dynamics on the model space must have frequency dependence.
    \item[(2)] For any choice of the 1-1 mapping $f$, if (1) is satisfied and $\langle \Psi_i | H_{ab} | \Psi_j \rangle = \langle f[\Psi_i] | H_{\mathrm{model}} | f[\Psi_j] \rangle$  $\forall \,\Psi_i,\Psi_j\in \mathcal{LE}$, the dynamics in the model system exactly match the dynamics of the \textit{ab initio} system.
    \item[(3)] $f$ can be constructed in any way, as long as it is 1-1 and (2) is true. However, some constructions of $f$ result in more complex effective Hamiltonians than others. 
    \item[(4)] Note also that $\mathcal{LE}$ must be the same dimension as $\mathcal{H}_{\text{model}}$; otherwise, a 1-1 mapping does not exist.
\end{enumerate}
Note that as long as the effective Hamiltonian satisfies (2), the dynamics are preserved, and therefore, the effective Hamiltonian is accurate. 
Poor choices of the mapping $f$ and the low-energy space $\mathcal{LE}$ result in complex $H_{\text{model}}$ with potentially many-body interaction terms in order to satisfy (2). 
Therefore, a useful downfolding process first identifies a low-energy space $\mathcal{LE}$ and a mapping $f$ such that only a few simple terms in $H_{\text{model}}$ are required.

In the following, we show some general strategies to select $\mathcal{LE}$ and $f$ for the \textit{ab initio} hydrogen chain, which applies to the strong coupling regime.
$\mathcal{LE}$ is selected via clustering in a descriptor space formed from relevant correlation functions, then the 1-1 mapping $f$ between the model and \textit{ab initio} eigenstates is constructed via matching their local descriptors, then finally, the effective model $H_{\text{model}}$ is determined by directly matching the \textit{ab initio} Hamiltonian operator to $H_{\text{model}}$. 
A comparison between R-DMD with other state-of-the-art downfolding methods is presented in the Discussion, where we demonstrate that R-DMD performs well both the atomic limit and away from the atomic limit due to its non-perturbative nature and the explicit renormalization of the spin operators.

\begin{figure}
    \centering
    \includegraphics{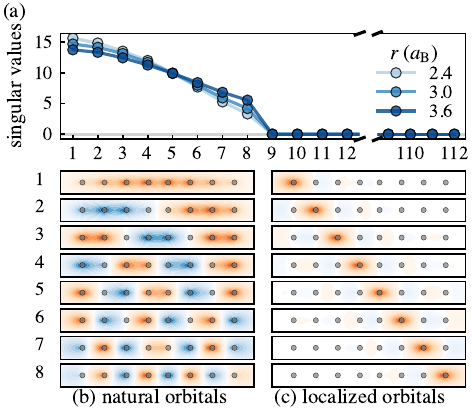}
    \caption{Determination of the minimal basis. (a). The singular values of the stacked 1-RDMs in the molecular orbital basis of the first 400 CASSCF eigenstates for $r = 2.4, 3.0, 3.6\,a_{\text{B}}$.
    (b). The natural orbitals, which are the column vectors corresponding to the eight non-zero singular values shown in (a), for $r = 3.0\,a_{\text{B}}$. 
    (c). The intrinsic atomic orbitals (IAOs) of the eight natural orbitals. These IAOs are localized in real space and used when computing the wave function descriptors.
    }
    \label{fig:results0_pca}
\end{figure}

\section{Renormalizaed density matrix downfolding}

In R-DMD, we aim to use high-accuracy \textit{ab initio} data to choose the low-energy space, construct the 1-1 mapping,  and fit the effective Hamiltonian. 
This technique is non-perturbative and based on high quality \textit{ab initio} solutions, and overcomes the issues of spectral fitting and unrenormalized density matrix downfolding (see section~\ref{section:other_methods}). 
Because the 1-1 map is constructed explicitly, R-DMD can also detect when a mapping is possible or not, thus allowing us to evaluate the appropriateness of a model, a capability that most downfolding methods lack.
A summary of the strategy is as follows.

\begin{enumerate}
    \item In this work, $\mathcal{LE}$ is selected based on clustering in a wave function descriptor space. This is done for the hydrogen chain case to correctly identify spin excitations versus charge excitations. A simple energy cutoff would not work. We believe this to be a powerful strategy that encompasses many options; for example, in graphene, one might wish to separate the $p_z$ bands from the others, which would be achieved by evaluating the occupation number on the $p_z$ states.
    \item The key innovation of R-DMD is to construct the 1-1 mapping $f$ not based on \textit{exact} similarity between model and \textit{ab initio} states, but approximate similarity. This is an intermediate between methods like spectral fitting and those like DMD or DFT+cRPA. Spectral fitting makes no guarantees that the physics of the model resembles the \textit{ab initio}. Methods such as DMD or DFT+cRPA, require the model to be written in terms of the bare electrons, resulting in complex Hamiltonians. This choice also improves the fitting process significantly versus spectral fitting. The effective model need not be solved, unlike spectral fitting.
    \item With $\mathcal{LE}$ chosen and $f$ constructed, the Hamiltonian parameters are simply fit to minimize the deviation between the model eigenenergies and the \textit{ab initio} eigenenergies. 
\end{enumerate}

{\onecolumngrid
\begin{center}
\begin{figure}
	\includegraphics{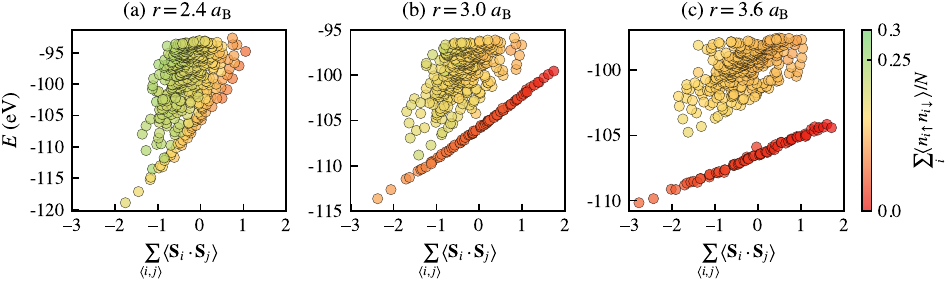}
    \caption{A low-energy space is assignable when there are distinguishable clusters of energy eigenstates in an extended descriptor space. Each point is an energy eigenstate of an 8-atom hydrogen chain. 
    The lines of states in which energy increases roughly linearly with $\sum_{\langle i,j\rangle}\langle \mathbf{S}_i \cdot \mathbf{S}_j \rangle$ can be mapped 1-1 to a spin Hilbert space. The statistical uncertainties are of order 0.01 for both quantities, and much smaller than the symbols. }
    \label{fig:clusters}
\end{figure}
\end{center}
}

\twocolumngrid 

\subsection{Constructing the \textit{ab initio} data set}
\label{subsection:descriptors}

 We use real-space variational Monte Carlo (VMC) to obtain approximate many-body wave functions for the eigenstates of the \textit{ab initio} Hamiltonian,
 \begin{equation}
	    H_{ab} =  \sum_{iI} \left(-\frac{1}{2}\nabla_{i}^{2}
	   -\frac{Z_I}{\left|\mathbf{r}_i - \mathbf{x}_{I}\right|}\right)
	   +\sum_{i \neq j} \frac{1}{\left|\mathbf{r}_{i}-\mathbf{r}_{j}\right|}.
	 \label{eqn:method0_ab_Hamiltonian}
	 \end{equation}
 The trial wave functions take a multi-Slater-Jastrow form,
 \begin{align}
	     \Psi_T(\mathbf{r}_1, \mathbf{r}_2, \cdots, \mathbf{r}_N)
	     &= e^{J(\mathbf{r}_1, \mathbf{r}_2, \cdots, \mathbf{r}_N)} \notag \\
      &\times \sum_k c_k D^{\uparrow}_k(\chi_{i \uparrow}(\mathbf{r}_j))D^{\downarrow}_k(\chi_{i \downarrow}(\mathbf{r}_j)),
	     \label{eqn:method1_vmc_trial_wf}
\end{align}
 where $\mathbf{r}_i$ and $\mathbf{x}_I$ represent the real-space coordinates of the $i$th electron and $I$th ion.
 The multi-Slater expansion $\sum_k c_k D^{\uparrow}_k(\chi_{i \uparrow}(\mathbf{r}_j))D^{\downarrow}_k(\chi_{i \downarrow}(\mathbf{r}_j))$, was generated using a multi-reference quantum chemistry method, restricted Hartree-Fock (RHF) + complete active space self-consistent field (CASSCF)~\cite{ruedenberg_are_1982}. 
 We started with RHF and optimized molecular orbitals variationally with a single Slater determinant as the ground state \textit{ansatz}.
 Then, in state-averaged CASSCF(8,$\,$(4,$\,$4)), we chose an active space that includes the first eight molecular orbitals and all the electrons, constrained in the $S_z = 0$ sector.
 The CASSCF \textit{ansatzes} for the eigenstates are expansions of determinants given by all the possible excitations within the active space.
 The molecular orbitals $\chi_{i\sigma}({\mathbf{r}})$ were parameterized using correlation consistent triple-zeta basis set \cite{dunning_gaussian_1989}, which includes up to the $3d$ orbitals of hydrogen atoms. 
 In the state-averaged CASSCF calculation, the coefficients $c_k$ and $\chi_{i\sigma}(\mathbf{r})$ are optimized to minimize the cost function, which is chosen as the energy averaged over the first $m$ eigenstates.
 Here, we chose $m=400$ based on the allowed computational resources.
 The trial wave functions for all the excited states contain the same fixed Jastrow function $e^{J}$ that includes electron-ion and electron-electron correlations, defined in Ref.~\cite{wheeler_pyqmc_2022}.
 The Jastrow function $J(\mathbf{r}_1, \cdots, \mathbf{r}_N)$ was optimized for the ground state wave function.
 We checked that this approximation does not change the derived model, but the model parameters can be improved with more accurate \textit{ab initio} wave functions.

The training data set for detecting the emergent low-energy Hilbert space $\mathcal{LE}$ is built with the following descriptors of the \textit{ab initio} eigenstates:
\begin{itemize}
	\item[-] the energy expectation value: 
	
	$E[\Psi_a]= \bra{\Psi_a} H_{\text{\textit{ab}}} \ket{\Psi_a}$,
	
	\item[-] the $l$th-nearest-neighbor hopping, $l=1,\,2,\,3$:
	
	$d_{t_{l}}[\Psi_a]=\sum_{i,\sigma}\bra{\Psi_a} c_{i \sigma}^{\dagger} c_{i+l \sigma}^{}+c_{i+l \sigma}^{\dagger} c_{i \sigma}^{}\ket{\Psi_a}$,
	
	\item[-] the total double occupancy:
	
	$d_{U}[\Psi_a]=\sum_{i}\bra{\Psi_a} n_{i \uparrow}n_{i \downarrow} \ket{\Psi_a},~ n_{i\sigma} = c^\dagger_{i\sigma}c_{i\sigma}^{}$,
	
	\item[-] the spin-spin correlation:
	
	$d_{J}[\Psi_a]=\sum_{\langle i, j\rangle}\bra{\Psi_a} \mathbf{S}_{i} \cdot \mathbf{S}_{j} \ket{\Psi_a}$,
\end{itemize}
where $c^\dagger_{i\sigma}$($c_{i\sigma}^{}$) creates(annihilates) an electron in orbital $\phi_{i\sigma}$ of the minimal basis set (see section~\ref{section:basis} for details in constructing the minimal basis set). 
$\mathbf{S}_i = \sum_{i, \alpha \alpha'}
c^\dagger_{i\alpha} \pmb{\sigma}_{\alpha\alpha'}^{}  c_{i\alpha'}^{}
$, where $\pmb{\sigma} = (\sigma_x, \sigma_y, \sigma_z)$ are the spin $1/2$ Pauli matrices.

\subsection{Finding a minimal basis set of a $\mathcal{LE}$}
\label{section:basis}

In general, constructing a localized basis set for a set of \textit{ab initio} eigenstates is not necessary --- one could optimize the basis when determining the $\mathcal{LE}$, enforcing the effective Hamiltonian to be as compact as possible.
However, in the case of the 8-atom hydrogen chain, we found using a localized basis set constructed in the following way sufficiently good for obtaining compact models in hydrogen chains.
Note that one might need to use generalized principal components of two-body or even higher-order reduced density matrices if one is to describe higher-order interactions.
Since the descriptors evaluated on the basis sets are only used for determining the 1-1 mapping $f$, as long as $f$ stays topologically the same, the choice of the basis sets does not matter.

For each interatomic separation $r$, we computed the principal components of the eigenstates' one-body reduced density matrices (1-RDMs) in the full molecular orbital basis.
Let us assume the total number of atomic orbitals which parameterize the many-body wave functions is $A$.
The spin $\sigma$ 1-RDM of eigenstate $\Psi_a$ in molecular orbital basis $\left\lbrace \chi_{i\sigma}\right\rbrace$ is given by
$\rho_{ij}^\sigma \left[\Psi_a\right] = \braket{\Psi_a | \tilde{c}^\dagger_{i\sigma} \tilde{c}_{j\sigma}^{} |\Psi_a}, \,\sigma = \uparrow,\,\downarrow$,
where $\tilde{c}^\dagger_{i\sigma}$ and $\tilde{c}_{i\sigma}$ are the creation and annihilation operators of an electron in molecular orbital $\chi_{i\sigma}$, expanded in terms of the $A$ atomic orbitals.
For each eigenstate, $\rho^{\sigma}$ is a matrix of size $A \times A$.
We included only the first 400 many-body eigenstates, which demonstrate our method well with affordable computational cost, in the \textit{ab initio} Hilbert space with ordering given by state-averaged CASSCF.
Then we stacked the 1-RDMs of both spin channels for all the eigenstates in columns and obtained an $A \times (800A)$ matrix $\varrho = 
\big(\rho^{\uparrow}[\Psi_1], \cdots, \rho^{\uparrow}[\Psi_{400}],
\rho^{\downarrow}[\Psi_1], \cdots, \rho^{\downarrow}[\Psi_{400}]
\big)$, and did singular value decomposition to $\varrho$: $\varrho = U\Sigma V$.
The column vectors of $U$ with non-zero singular values are the natural orbitals of the Hilbert space spanned by the first 400 eigenstates.
Lastly, we computed the intrinsic atomic orbitals (IAOs) \cite{knizia_intrinsic_2013} of the natural orbitals. 
These IAOs are localized in real space and were used for building the \textit{ab initio} data set (see Fig.~\ref{fig:results0_pca} for details).

\subsection{Selection of $\mathcal{LE}$ via clustering}

In Fig~\ref{fig:clusters}, we show the results of the \textit{ab initio} solutions of the 8-atom hydrogen chain. 
Details of the \textit{ab initio} calculations can be found in section~\ref{subsection:descriptors}.
Starting from the large separation distance of $r=3.6\,a_{\mathrm{B}}$ (panel (c)), one can easily identify spin-like excitations, in which the double occupancy is low, and the energy varies approximately linearly with the spin correlation $d_{J}$, as one would expect for a Heisenberg-like model. 
At higher energies, one can see the charge excitations, which have significant double occupancy. 
As the separation is decreased, the double occupancy in the spin excitations increases, and the charge excitations are closer to energy; however, at $r=3.0\,a_{\mathrm{B}}$, the two groups are still visible by eye.
Finally, by $r=2.4\,a_{\mathrm{B}}$, still before the metal-insulator transition, the charge and spin excitations have merged and are not distinguishable except for a few states at very low energy.

\begin{figure}
	\includegraphics{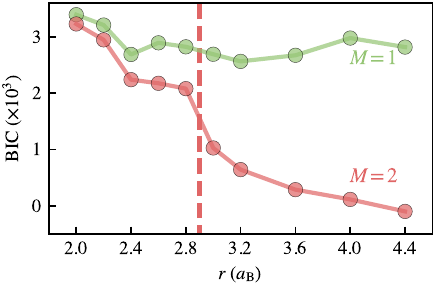}
	\caption{Detection of clusters using a Gaussian mixture model. Above $2.9\,a_{\mathrm{B}}$, there is a significant improvement in the Bayesian information criterion (BIC) for two clusters ($M=2$), indicating the presence of two classes of eigenstate.}
	\label{fig:gmm_bic}
\end{figure}

An obvious choice for $\mathcal{LE}$ is to use an energy cutoff, as is common in the renormalization group. 
A problem with this choice is demonstrated in Fig~\ref{fig:clusters}(b) and (c), in which there are clear groups of eigenstates when the spin-spin, double occupancy, and energy are considered, but these states overlap in energy. 
Only when the charge excitations merge with the spin excitations are there not distinguishable spin and charge sectors. 
While this can be seen by eye for a simple system like this, to systematically classify the eigenstates, we used a Gaussian mixture model (GMM).
The GMM is a probabilistic model that assumes a set of descriptors $\mathbf{d}$ of size $n$ is described by a mixture of $M$ Gaussian distributions \cite{reynolds_gaussian_2009},
In particular, it allows for the clusters to be highly anisotropic, as they are in the case here.
In this work, we used the machine learning package scikit-learn \cite{pedregosa_scikit-learn_2011} to determine the distribution with full rank covariance matrices. 
In the case of the hydrogen chain, $\mathbf{d}$ includes all the descriptors listed in section~\ref{subsection:descriptors}.
We find that the energy, double occupancy, and spin correlations are the most relevant descriptors; adding more does not change the results of the clustering algorithm.

In order to compute the number of optimal clusters of eigenstates, we chose to use the Bayesian information criterion (BIC) to measure the accuracy of the GMM \cite{schwarz_estimating_1978} and apply the heuristic elbow criterion \cite{kodinariya_review_2013} to determine $M$ in the GMM when clustering the eigenstates.
The BIC score penalizes the sample size and the total number of parameters to avoid overfitting, so that we are able to detect the emergence of disjoint eigenstate clusters that can be assigned to a low-energy space.
With this method, $\mathcal{LE}$ is not dictated by an energy cutoff, but instead, an extended descriptor space (energy, spin-spin correlation, and double occupancy) is necessary for it to be detected.

Fig.~\ref{fig:gmm_bic} shows the BIC scores of one- and two-cluster GMMs, for hydrogen chains with interatomic separations $r\geq 2.0\,a_{\mathrm{B}}$, obtained using all the descriptors available.
The decrease in BIC for a two-cluster GMM at $3.0\,a_{\mathrm{B}}$ clearly shows that the \textit{ab initio} eigenstates can be assigned to two groups (the spin- and charge-like excitations).
According to this metric, the spin excitations merge into the charge excitations at around $3.0\,a_{\text{B}}$, i.e., an interatomic separation that is unrelated to the ground state metal-insulator transition, indicating the formation of a complex correlated state without a well-separated magnetic spectrum.
Note that distinct spin and charge sectors exist even when the double occupancy of the ground state is quite large; at $3.0\,a_{\mathrm{B}}$, it is almost half the noninteracting value of 0.25 in the ground state.

\begin{figure}
    \centering
    \includegraphics{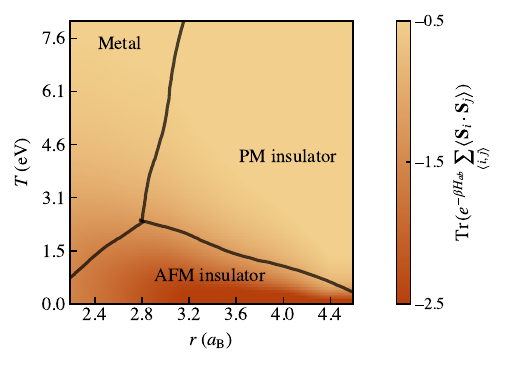}
    \caption{A sketch of the phase diagram for the hydrogen chain, based on the calculated charge fluctuation and the spin-spin correlation (see appendix~\ref{appendix:phase_diagram_additional_info}).
    The colors represent the thermal expectation values of the spin-spin correlations, computed using the \textit{ab initio} eigenstates of the 8-atom chain. 
    The phase boundaries were sketched based on the values of charge fluctuations and spin-spin correlations (see Fig.~\ref{fig:appendix1_phase_diagram_additional_info} for details).
    }
    \label{fig:phase_diagram}
\end{figure}

We will note here that the selection method is ultimately justified by the quality of $H_{\text{eff}}$. 
It is an interesting avenue of future research for more complex systems, whether clusters represent the best partitioning or whether there are more effective methods. 
The key contribution of this work is that we now have a systematic way of internally assessing one method versus another for any system.

The presence or absence of distinguishable clusters correlates with physics in this system. 
In Fig.~\ref{fig:phase_diagram}, an approximate phase diagram is shown. 
The point at which the clusters become indistinct ($2.8\,a_{\mathrm{B}}$) corresponds to the point at which the paramagnetic insulator phase disappears in the phase diagram, and a metallic antiferromagnet appears, which is a direct indication of the transition from a Slater insulator to a Mott insulator.
In Slater insulators, the insulating ground state arises from the antiferromagnetic (AFM) order.
Therefore, all the excitations that change the spin would also involve finite charge excitations.
This can be seen from Fig.~2(a) (at $r=2.4~a_{\text{B}}$), where the color of the markers corresponds to the double occupancy per site which measures the amount of charge excitations.
As for the Mott insulator scenario, which correspond to the larger $r$ (see Fig.~2(b-c)), there exist excitations that change spin but do not involve any change in the double occupancies (i.e., the spin-like excitations).
The separation between these two regimes is precisely whether we have a clear spin/charge separation or not, i.e., whether there is clustering of the eigenstates in the descriptor space.

\subsection{Construction of the 1-1 mapping $f$}

\begin{figure}
	\includegraphics[width=0.5\textwidth]{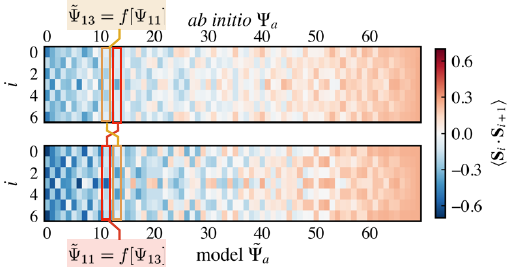}
    \caption{The optimal 1-1 mapping $\tilde{\Psi} = f[\Psi]$ between the ab-initio and model eigenstates: the local spin correlation of the many-body eigenstates at $r=3.0\,a_{\text{B}}$.
    The model eigenstates are eigenstates of the Heisenberg model.
    States (columns) are labeled according to their energies and matched according to the similarity of the local descriptors (rows).
    For example, $\Psi_{11}$ is mapped to model state $\tilde{\Psi}_{13}$ (highlighted in yellow boxes), and $\Psi_{13}$ is mapped to $\tilde{\Psi}_{11}$ (red boxes).
   }
    \label{fig:construct_f}
\end{figure}

Our ultimate criterion is that the \textit{ab initio} Hamiltonian operator is reproduced in the model, i.e., $H_{\text{eff}}$ satisfies
\begin{equation}
    \langle \Psi_i | H_{ab} | \Psi_j \rangle = \langle f[\Psi_i] | H_{\mathrm{eff}} | f[\Psi_j] \rangle,
    \label{eqn:hamiltonian_matching}
\end{equation}
for all $\Psi_i, \Psi_j \in \mathcal{LE}$, then the dynamics in the model system are the same as the \textit{ab initio}. 
This requires a functional mapping $f$ between the indices of the two operators. 
$f$ can be chosen arbitrarily so long as Eqn~\ref{eqn:hamiltonian_matching} is satisfied. 
Therefore, one can choose any heuristic, so long as Eqn~\ref{eqn:hamiltonian_matching} can be matched.

In the case of mapping the hydrogen chain to a spin model, we take advantage of a simplification that occurs since we are interested in finding the best Heisenberg model. 
The energy eigenstates of the Heisenberg model are identical, independent of the value of $J$. 
We thus associate the \textit{ab initio} eigenstate $\Psi_a$ with the model eigenstate $\tilde{\Psi}_b$ based on a similarity of the local spin-spin correlations. 
An example is shown in Fig~\ref{fig:construct_f}. 
We construct a vector of descriptors in first principles $d_{ai} = \langle \Psi_a | \mathbf{S}_i \cdot \mathbf{S}_{i+1} | \Psi_a \rangle$ and the equivalent for the model space. 
Then we compute the $N\times N$ distance matrix $D_{ab} = \sum_i (d_{ai} - \tilde{d}_{bi})^2$. 
The assignment problem is then solved using the linear assignment method~\cite{kuhn_hungarian_1955}.

 In more complex situations, this method can be extended as follows.
Define
\begin{equation}
	D_{ij}  = \langle \Psi_i | \hat{D} | \Psi_j \rangle
\end{equation}
and 
\begin{equation}
	\tilde{D}_{ij} = \langle \tilde{\Phi}_i | \hat{\tilde{D}} | \tilde{\Phi}_j \rangle,
\end{equation}
for operators $\hat{D}$ in the ab initio space, corresponding operator $\hat{\tilde{D}}$ in the model space, and $\tilde{\Phi}_i$ some orthogonal basis in the model space. 
Then, the optimal assignment can be given by choosing a rotation matrix $R$ such that the objective function 
\begin{equation}
	\lVert D - R^T \tilde{D} R \rVert ^2
\end{equation}
is minimized. 
Our simpler strategy above is equivalent to this with the constraint that the rotation matrix can only interchange eigenstates of the Heisenberg model, not create superpositions of them.

We will also note here that $f$ could be assigned in other ways; it is an interesting research avenue to explore what heuristics result in the most accurate and simple $H_{\text{eff}}$ for common systems.
One could potentially apply machine learning tools to select the correct assignment of states for a given criterion of simplicity and accuracy of $H_{\text{eff}}$.
As we shall show later, the method we use here results in accurate Heisenberg models for the hydrogen chain.
We expect, since it is physically motivated, methods such as the one presented here will be useful.

\begin{figure}
    \centering    
    \includegraphics{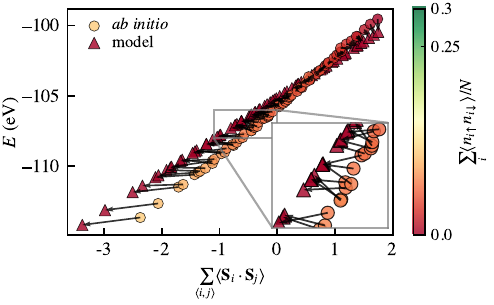}
    \caption{Optimal Heisenberg model with mapping determined from Fig~\ref{fig:construct_f} with $r=3.0\,a_{\mathrm{B}}$. 
    The energy of the model states is computed with the optimal Heisenberg model. 
    A perfect mapping would have the model and \textit{ab initio} states with exactly the same energy.
    The mapping renormalizes the spin operator explicitly for each wave function in $\mathcal{LE}$; the horizontal distance is the renormalization of spin.
    In the figure, the black arrows indicate this state-wise renormalization of the spin operator.
    }
    \label{fig:assignment}
\end{figure}

The result of the assignment is shown in Fig~\ref{fig:assignment}. 
At the interatomic separation of 3.0\,$a_{\text{B}}$, the effective spins in the model are quite different from the \textit{ab initio} spins.
From this assignment, one can see the effects of renormalization on the model description of the antiferromagnetic ground state, and very little in the ferromagnetic highest energy state, since the mapped model spin correlations are very different for the antiferromagnetic state, while very similar for the ferromagnetic state. 
The mapping is not a constant renormalization of the magnitude of the spin operators; on the contrary, the spin moment measured in the model space is related to the \textit{ab initio} spin by a state-dependent factor. 
Without constructing the mapping explicitly, it would be very difficult to guess the spin renormalization, even in this simple case.

\subsection{Determination of $H_{\mathrm{eff}}$}

\begin{figure}
    \includegraphics[]{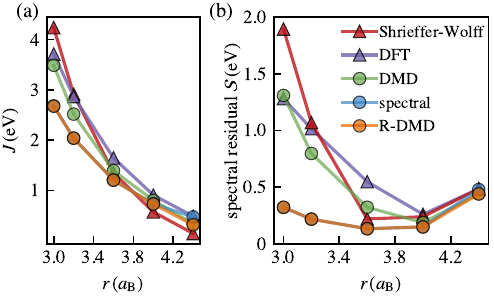}
    \caption{(a) The derived $J$ values and (b) the spectral residual (error in the energy eigenvalues) evaluated using Schrieffer-Wolff transformation, DFT, DMD, spectral-fitting DMD (spectral), and R-DMD versus the interatomic separation.
    R-DMD obtains accurate models for the strong coupling regime $r\leq 3.6\,a_{\mathrm{B}}$, in contrast to standard methods.
    }
    \label{fig:accuracy}
\end{figure}

If $H_{\text{eff}}$ satisfies Eqn~\ref{eqn:hamiltonian_matching}, then the dynamics in the model system are the same as that given by \textit{ab initio} calculations. 
As $\ket{\Psi} \in \mathcal{LE}$ since $\ket{\Psi} = \sum_{a} \alpha_{a} \ket{\Psi_a}$.
Similarly $\ket{f[\Psi]}$ is in the model Hilbert space since $\ket{f[\Psi]} = \sum_{a} \alpha_{a} \ket{f[\Psi_a]}$ where $f[\Psi_a]$ is an eigenstate of $H_{\text{eff}}$. 
With there being a 1-1 mapping of $\ket{\Psi_{a}}$ to $\ket{f[\Psi_{a}]}$ the expression for the dynamics below follows,
\begin{equation}
    \exp(-i H_{ab} t) |\Psi\rangle = f^{-1}\left[\exp(-i H_{\mathrm{eff}} t) |f[\Psi]\rangle \right].
\end{equation}
Because we are matching energy eigenstates to energy eigenstates, the off-diagonal elements of Eqn~\ref{eqn:hamiltonian_matching} are automatically satisfied. 
We thus fit $J$ and $E_0$ to minimize the loss function 
\begin{equation}
    \mathcal{S} = \sum_a\left( J\langle f[\Psi_a]| \tilde{\mathbf{S}}_i \cdot \tilde{\mathbf{S}}_j | f[\Psi_a] \rangle + E_0 - \langle \Psi_a | H_{ab} | \Psi_a \rangle \right)^2,
    \label{equation:spectral_residual}
\end{equation}
where $\tilde{\mathbf{S}}_i$ are spin-1/2 operators in the model spin Hilbert space.

\section{Comparison with other downfolding techniques}

\label{section:other_methods}

\begin{table*}
\centering
	\caption{A comparison of common approaches to downfolding to low energy models. }
	\label{table:downfolding_comparison}
	\begin{tabular}{p{1.7in}p{1.2in}p{1.5in}p{2.3in}}
		Method & $\mathcal{LE}$ choice & $f$ & $H_{\text{eff}}$\\
		\hline
		\hline
		DFT+cRPA & Active orbitals & Bare & $H_{\text{KS}} + H_{\text{DC}} + H_{\text{cRPA}}$ \\
		\hline
		Shrieffer-Wolff transformation & Eigenstates of $H_0$ &  & Perturbatively renormalized\\
		\hline
		Canonical transformation & Rotation & Bare & In principle exact, but usually truncated \\
		\hline
		Spectral fitting &  Eigenstates of $H_{ab}$ & Closest match of spectrum & Best fit to \textit{ab initio} data\\
		\hline
	    DMD &  Eigenstates of $H_{ab}$ & Bare & Best fit to \textit{ab initio} data\ \\
		\hline
		\multirow{2}{*}{R-DMD} &  \multirow{2}{*}{Eigenstates of $H_{ab}$} &  Match of both & \multirow{2}{*}{Best fit to \textit{ab initio} data} \\
        & & spectrum \& eigenstates & \\
		\hline
	\end{tabular}
\end{table*}

In this section, we first introduce other state-of-the-art downfolding techniques, including symmetry-broken DFT, DFT+cRPA, Schrieffer-Wolff transformation, spectral fitting, and unrenormalized density matrix downfolding (DMD).
We then compare the results given by R-DMD with these methods.

\textbf{Broken-symmetry DFT:}
In this technique~\cite{shoji_general_2006}, two or several broken-symmetry magnetic states are constructed and then variationally obtained via DFT. 
The parameters in $H_{\mathrm{eff}}$ are determined by matching the DFT energy difference between, for example, the AFM and FM states with that predicted by a spin model.
This method can be applied to systematically derive exchange interactions from DFT.
To compare with R-DMD, we evaluated the Heisenberg exchange from DFT as $E_{\text{FM}}-E_{\text{N\'eel}}$ in a 2-atom cell hydrogen chain with periodic boundary condition with 10 $k$ points.
The results are presented as purple triangles in Fig.~\ref{fig:accuracy}.

\textbf{DFT+cRPA:} In this technique~\cite{aryasetiawan_frequency-dependent_2004}, a set of correlated one-particle orbitals (``active orbitals") is chosen, and the screened Coulomb interaction between them is evaluated.
Because $\mathcal{LE}$ is chosen based on active orbitals, it is not closed under the operation of the Hamiltonian, and a frequency-dependent interaction emerges. 
For the problem considered here (downfolding onto a spin model), a straightforward application of DFT+cRPA is elusive.

\textbf{Schrieffer-Wolff transformation}~\cite{schrieffer_relation_1966, bravyi_schriefferwolff_2011}
is a venerable perturbative method that requires that the projected subspace be energetically separated from the rest of the Hilbert space. 
For this reason, for the case of the hydrogen chain to spin model mapping, this method is expected to work well for large interatomic separations.
As we shall show, it is possible to map the low energy spectrum of the hydrogen chain onto a spin model well outside the perturbative regime.
To compare with R-DMD, we evaluated the Heisenberg exchange from Schrieffer-Wolff transformation using the $t_1$ and $U$ given by downfolding to a Hubbard model, see appendix~\ref{appendix:hubbard}.
The results are presented as red triangles in Fig.~\ref{fig:accuracy}.

\textbf{Spectral fitting}
is a technique that is similar to R-DMD, in that it uses reference \textit{ab initio} calculations. 
In this method, a model is assumed, and the parameters are varied to best represent the eigenenergies of the model.
A common application of this method is the derivation of the hopping parameters in tight-binding models without wannierization of the Kohn-Sham orbitals, but simply through fitting to either experiment~\cite{kuzmenko_determination_2009} or Kohn-Sham spectra~\cite{ribeiro_strained_2009}.
In contrast to cRPA and Schrieffer-Wolff, such a method is nonperturbative; however, it requires eigenenergies to be matched to one another. 
As we have shown for the \textit{ab initio} hydrogen chain, it is not necessarily the case that states with the same energy actually correspond to states with similar properties, so a criterion based solely on energy may scramble the physics.

\textbf{Unrenormalized density matrix downfolding} (DMD)~\cite{changlani_density-matrix_2015,zheng_real_2018, chang_effective_2020} is a method that attempts to overcome some of the issues with spectral fitting regarding the properties of the wave function, and also does not require one to obtain exact eigenstates. 
In this method, one matches the expectation value of the Hamiltonian between the model and \textit{ab initio} states. 
However, it does not ensure a 1-1 mapping between the two spaces.
It is therefore named ``unrenormalized'' since the operator values used to derive the model are computed using the bare expression of the single-particle basis.

\vskip 1cm

In Fig~\ref{fig:accuracy}, we compare the derived $J$ and the error compared to the underlying \textit{ab initio} data as a function of interatomic spacing. 
We can see that R-DMD performs well in terms of obtaining the minimal spectral residual (Eqn~\ref{equation:spectral_residual}) among all methods, comparable with spectral fitting DMD (spectral). 
We compare to the DFT-computed $J$ from broken-symmetry magnetic states, the Schreiffer-Wolf perturbative calculation from an effective $t-U$ model (see appendix~\ref{appendix:hubbard}), density matrix downfolding (DMD), and matching the spectrum without matching the properties of the states (spectral). 
Perturbative methods such as Schreiffer-Wolf work well for large interatomic separation $r$, but overestimate $J$ at smaller $r$. 
Unlike R-DMD, perturbative techniques also give no indication of their accuracy/applicability at lower interatomic separations. 
Broken symmetry DFT calculations are similar, both in the lack of indication of applicability and in the poor calculation of the $J$ term at lower $r$ due to inaccuracies in the density functional.
In this simple case, spectral fitting DMD is very similar to R-DMD, where the \textit{ab initio} eigenstates are matched to the model eigenstates based on the energy ordering. 
However, in the general case, spectral fitting would require many diagonalizations of the effective Hamiltonian, while R-DMD does not require the effective Hamiltonian to be diagonalized at all.

\section{Conclusions}

We have presented a new approach to the determination of multiscale models, called renormalized density matrix downfolding (R-DMD), bringing the derivation of electronic models to the same footing as atomic force field models. 
In the \textit{ab initio} hydrogen chain, using the largest-scale and high-accuracy QMC calculations for up to frequency ranges about 10\,eV, we have demonstrated that R-DMD performs well both at the atomic limit and away from the atomic limit due to its non-perturbative nature and explicit renormalization of the \textit{ab initio} operators.
This approach is systematically improvable in two senses: the underlying data can be improved systematically by considering progressively more accurate solutions of the \textit{ab initio} problem, and the effective Hilbert space and Hamiltonian can be systematically improved to better represent the underlying \textit{ab initio} data.  
Note that this method does not require exact energy eigenstates, so long as the ab initio data is a superposition of the eigenstates that span $\mathcal{LE}$, nor does it require the model Hamiltonian to be diagonalized.
Renormalization of quantum operators is determined as a byproduct of the procedure and is explicitly constructed. 

R-DMD is a flexible framework in which the mapping function $f$ and the choice of the low-energy space $\mathcal{LE}$ can be selected at will to produce the most accurate and simple resultant Hilbert space and Hamiltonian. 
Eqn~\ref{eqn:hamiltonian_matching} gives a variational criterion for constructing low-energy models. The selection of $\mathcal{LE}$ and $f$ are arbitrary as long as $\mathcal{LE}$ is spanned by eigenstates and $f$ is 1-1. 
It is also possible to adapt R-DMD to treat systems in which $\mathcal{LE}$ is not closed under the application of $H_{ab}$. 
In this case, the energy-conserving Hamiltonian framework used here can be simply extended to a Lagrangian or Linbladian description in the model space; the principle of matching operators still applies.

\section{Acknowledgements}
 The initial contributions of Y.C. in performing calculations, generating the data, and developing the theory, and the contributions of L.K.W. in supervising, writing, and creating the theory were supported by the U.S. Department of Energy, Office of Science, Office of Basic Energy Sciences, Computational Materials Sciences Program, under Award No. DE-SC0020177.
 The contribution of Y.C. in writing was funded by the Abrahams Postdoctoral Fellowship from the Center for Materials Theory, Department of Physics \& Astronomy at Rutgers University.
The contributions of S.J. in developing the method of systematically constructing the 1-1 mapping $f$ and writing were supported by the National Science Foundation under Grant No. DGE-1922758.
This work made use of the Illinois Campus Cluster, a computing resource that is operated by the Illinois Campus Cluster Program (ICCP) in conjunction with the National Center for Supercomputing Applications (NCSA) and which is supported by funds from the University of Illinois at Urbana-Champaign.
This research used resources of the Oak Ridge Leadership Computing Facility at the Oak Ridge National Laboratory, which is supported by the Office of Science of the U.S. Department of Energy under Contract No. DE-AC05-00OR22725.

\appendix
\renewcommand\thefigure{A\arabic{figure}} 
\setcounter{figure}{0} 

\section{Clustering of eigenstates in the descriptor space}
\label{appendix:phase_diagram_additional_info}

\begin{figure}
\centering
\includegraphics{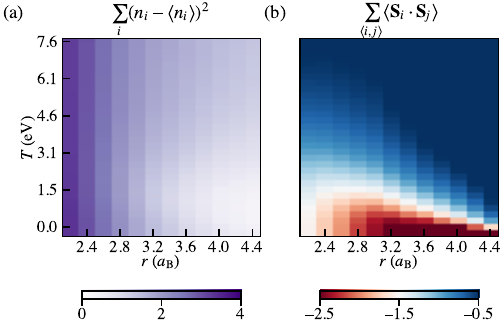}
\caption{(a) Computed charge fluctuation ($\sum_i \braket{(n_i - \braket{n_i})^2}$) and (b) spin-spin correlation ($\sum_{\braket{i,j}} \braket{\mathbf{S}_i\cdot\mathbf{S}_j}$) using the VMC eigenstates for atomic separations $r\geq 2.2 \,a_{\text{B}}$ at different temperatures..}
\label{fig:appendix1_phase_diagram_additional_info}
\end{figure}

Figure~\ref{fig:appendix1_phase_diagram_additional_info} shows the charge fluctuation ($\sum_i \braket{(n_i - \braket{n_i})^2}$) and the spin-spin correlation ($\sum_{\braket{i,j}} \braket{\mathbf{S}_i\cdot\mathbf{S}_j}$) as a function of the bond length and the temperature as given by the VMC eigenstates.
Note that the ground state of the system when $r\!\geq\! 2.2\,a_{\text{B}}$ is always an AFM insulator. 
A metallic state is characterized by its large charge fluctuation.
The phase boundaries in Fig.~5 are only schematic as given approximately by the contour lines that correspond to charge fluctuation $\!=\!2.0$ and spin-spin correlation $\!=\!-1.5$.

\section{Hubbard model derived using matching pursuit} 
\label{appendix:hubbard}

\begin{figure*}
\centering
\includegraphics{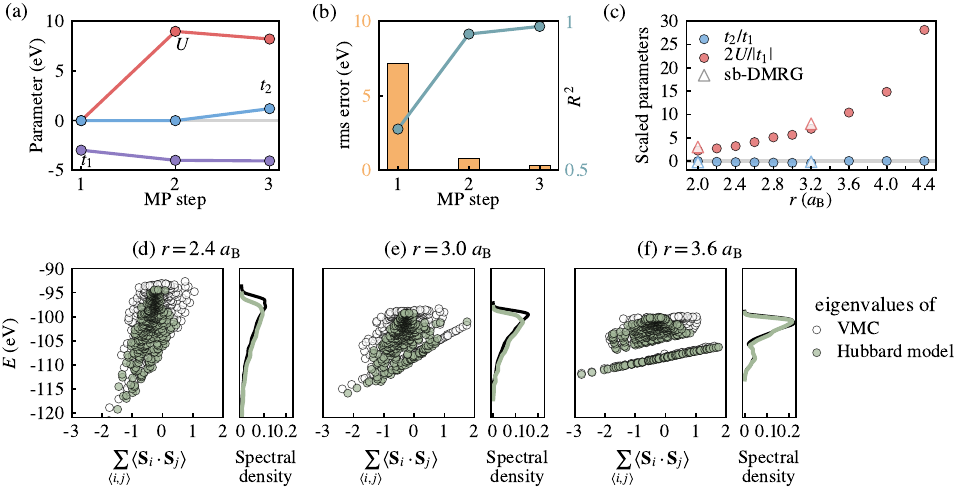}
\caption{(a) Parameter values for each matching pursuit (MP) step. 
A zero value indicates that the parameter is not yet included in the fit. 
(b) RMS error of the effective model and $R^2$ of the fit at each MP step. 
(c) Scaled parameter values ($t_2/t_1$ and $2U/|t_1|$) for each interatomic separation. 
For $r\geq 3.6\,a_{\text{B}}$, a $t_1-U$ model is chosen as the optimal model, so $t_2=0$. 
The results of the model derived using sliced-basis DMRG (sb-DMRG, see Ref.~\cite{sawaya_constructing_2022}) are plotted in comparison.
(d-f) The VMC (multi-Slater-Jastrow wave functions with fixed Jastrow) eigenvalues and Hubbard model eigenvalues for $r=2.4,\,3.0,\,3.6\,a_{\text{B}}$, along with the comparison between their spectral densities.
}
\label{fig:appendix2_hubbard_summary}
\end{figure*}

In the paper, we described how the AFM Heisenberg model is detected from highly accurate \textit{ab initio} many-body calculations. 
We discovered that the Heisenberg model only emerges for certain interatomic separations when the \textit{ab initio} eigenstates can be clustered into two sub-clusters.
In fact, for hydrogen chains with AFM insulating ground states, the Hubbard model well describes all the eigenstates of interest, including both the spin and charge excitations.
Figure~\ref{fig:appendix2_hubbard_summary} (a) shows the parameter values when we use the matching pursuit (MP) algorithm \cite{mallat_matching_1993} to select the important terms in the effective Hamiltonian.
MP chooses the descriptors based on their correlation with the residual of the previous fit. 
At MP step 1, we fit the \textit{ab initio} data to a $t_1$ only model. 
Then, at MP step 2, $U$ is included in the effective Hamiltonian since it correlated the most with the residual of a $t_1$ mode. 
At MP step 3, $t_2$ is then included.
The fit is improved as more descriptors are included in the effective Hamiltonian (see Fig.~\ref{fig:appendix2_hubbard_summary} (b)), as the RMS error is reduced and $R^2$ is increased.
In Fig.~\ref{fig:appendix2_hubbard_summary} (c), we summarized the scaled fitted parameters as a function of the interatomic separation $r$ and compared them with the state-of-the-art downfolded results using sliced-basis DMRG (sb-DMRG) \cite{sawaya_constructing_2022}.
Our results agree with the sb-DMRG values.

In Fig.~\ref{fig:appendix2_hubbard_summary} (d-f) panels, we compared the \textit{ab initio} eigenvalues and the eigenvalues yielded by diagonalizing the fitted Hubbard models, along with their spectral densities.
The distribution of the eigenvalues roughly agrees, while we note that there are discrepancies at the high energy.
We think this is due to two possible reasons.
At higher energy, the next-nearest-neighbor interaction is relevant. 
However, due to the lack of higher-energy \textit{ab initio} data, we were not able to capture the effect of next-nearest-neighbor interaction.
Also, similar to the Heisenberg model, the Hubbard model also requires renormalization using quasi-particle operators to accurately reproduce the \textit{ab initio} properties. 
In Fig.~\ref{fig:accuracy}, for the Schrieffer-Wolff transformation results, the $J$ values are computed as $J = \frac{4t_1^2}{U}$, using the $t_1$ and $U$ parameters obtained as above.
The spectral residual of the Schrieffer-Wolff model is then evaluated with a fitted intersection $E_0$ value by minimizing the spectral residual.


\begin{thebibliography}{0}%
\makeatletter
\providecommand \@ifxundefined [1]{%
 \@ifx{#1\undefined}
}%
\providecommand \@ifnum [1]{%
 \ifnum #1\expandafter \@firstoftwo
 \else \expandafter \@secondoftwo
 \fi
}%
\providecommand \@ifx [1]{%
 \ifx #1\expandafter \@firstoftwo
 \else \expandafter \@secondoftwo
 \fi
}%
\providecommand \natexlab [1]{#1}%
\providecommand \enquote  [1]{``#1''}%
\providecommand \bibnamefont  [1]{#1}%
\providecommand \bibfnamefont [1]{#1}%
\providecommand \citenamefont [1]{#1}%
\providecommand \href@noop [0]{\@secondoftwo}%
\providecommand \href [0]{\begingroup \@sanitize@url \@href}%
\providecommand \@href[1]{\@@startlink{#1}\@@href}%
\providecommand \@@href[1]{\endgroup#1\@@endlink}%
\providecommand \@sanitize@url [0]{\catcode `\\12\catcode `\$12\catcode
  `\&12\catcode `\#12\catcode `\^12\catcode `\_12\catcode `\%12\relax}%
\providecommand \@@startlink[1]{}%
\providecommand \@@endlink[0]{}%
\providecommand \url  [0]{\begingroup\@sanitize@url \@url }%
\providecommand \@url [1]{\endgroup\@href {#1}{\urlprefix }}%
\providecommand \urlprefix  [0]{URL }%
\providecommand \Eprint [0]{\href }%
\providecommand \doibase [0]{https://doi.org/}%
\providecommand \selectlanguage [0]{\@gobble}%
\providecommand \bibinfo  [0]{\@secondoftwo}%
\providecommand \bibfield  [0]{\@secondoftwo}%
\providecommand \translation [1]{[#1]}%
\providecommand \BibitemOpen [0]{}%
\providecommand \bibitemStop [0]{}%
\providecommand \bibitemNoStop [0]{.\EOS\space}%
\providecommand \EOS [0]{\spacefactor3000\relax}%
\providecommand \BibitemShut  [1]{\csname bibitem#1\endcsname}%
\let\auto@bib@innerbib\@empty
\end{thebibliography}%


\begin{thebibliography}{70}%
\makeatletter
\providecommand \@ifxundefined [1]{%
 \@ifx{#1\undefined}
}%
\providecommand \@ifnum [1]{%
 \ifnum #1\expandafter \@firstoftwo
 \else \expandafter \@secondoftwo
 \fi
}%
\providecommand \@ifx [1]{%
 \ifx #1\expandafter \@firstoftwo
 \else \expandafter \@secondoftwo
 \fi
}%
\providecommand \natexlab [1]{#1}%
\providecommand \enquote  [1]{``#1''}%
\providecommand \bibnamefont  [1]{#1}%
\providecommand \bibfnamefont [1]{#1}%
\providecommand \citenamefont [1]{#1}%
\providecommand \href@noop [0]{\@secondoftwo}%
\providecommand \href [0]{\begingroup \@sanitize@url \@href}%
\providecommand \@href[1]{\@@startlink{#1}\@@href}%
\providecommand \@@href[1]{\endgroup#1\@@endlink}%
\providecommand \@sanitize@url [0]{\catcode `\\12\catcode `\$12\catcode
  `\&12\catcode `\#12\catcode `\^12\catcode `\_12\catcode `\%12\relax}%
\providecommand \@@startlink[1]{}%
\providecommand \@@endlink[0]{}%
\providecommand \url  [0]{\begingroup\@sanitize@url \@url }%
\providecommand \@url [1]{\endgroup\@href {#1}{\urlprefix }}%
\providecommand \urlprefix  [0]{URL }%
\providecommand \Eprint [0]{\href }%
\providecommand \doibase [0]{https://doi.org/}%
\providecommand \selectlanguage [0]{\@gobble}%
\providecommand \bibinfo  [0]{\@secondoftwo}%
\providecommand \bibfield  [0]{\@secondoftwo}%
\providecommand \translation [1]{[#1]}%
\providecommand \BibitemOpen [0]{}%
\providecommand \bibitemStop [0]{}%
\providecommand \bibitemNoStop [0]{.\EOS\space}%
\providecommand \EOS [0]{\spacefactor3000\relax}%
\providecommand \BibitemShut  [1]{\csname bibitem#1\endcsname}%
\let\auto@bib@innerbib\@empty
\bibitem [{\citenamefont {G\l{}azek}\ and\ \citenamefont
  {Wilson}(1993)}]{glazek_renormalization_1993}%
  \BibitemOpen
  \bibfield  {author} {\bibinfo {author} {\bibfnamefont {S.~D.}\ \bibnamefont
  {G\l{}azek}}\ and\ \bibinfo {author} {\bibfnamefont {K.~G.}\ \bibnamefont
  {Wilson}},\ }\bibfield  {title} {\bibinfo {title} {Renormalization of
  {Hamiltonians}},\ }\href {https://doi.org/10.1103/PhysRevD.48.5863}
  {\bibfield  {journal} {\bibinfo  {journal} {Phys. Rev. D}\ }\textbf {\bibinfo
  {volume} {48}},\ \bibinfo {pages} {5863} (\bibinfo {year}
  {1993})}\BibitemShut {NoStop}%
\bibitem [{\citenamefont {G\l{}azek}\ and\ \citenamefont
  {Wilson}(1994)}]{glazek_perturbative_1994}%
  \BibitemOpen
  \bibfield  {author} {\bibinfo {author} {\bibfnamefont {S.~D.}\ \bibnamefont
  {G\l{}azek}}\ and\ \bibinfo {author} {\bibfnamefont {K.~G.}\ \bibnamefont
  {Wilson}},\ }\bibfield  {title} {\bibinfo {title} {Perturbative
  renormalization group for {Hamiltonians}},\ }\href
  {https://doi.org/10.1103/PhysRevD.49.4214} {\bibfield  {journal} {\bibinfo
  {journal} {Phys. Rev. D}\ }\textbf {\bibinfo {volume} {49}},\ \bibinfo
  {pages} {4214} (\bibinfo {year} {1994})}\BibitemShut {NoStop}%
\bibitem [{\citenamefont {Wilson}(1975)}]{RevModPhys.47.773}%
  \BibitemOpen
  \bibfield  {author} {\bibinfo {author} {\bibfnamefont {K.~G.}\ \bibnamefont
  {Wilson}},\ }\bibfield  {title} {\bibinfo {title} {The renormalization group:
  Critical phenomena and the {Kondo} problem},\ }\href
  {https://doi.org/10.1103/RevModPhys.47.773} {\bibfield  {journal} {\bibinfo
  {journal} {Rev. Mod. Phys.}\ }\textbf {\bibinfo {volume} {47}},\ \bibinfo
  {pages} {773} (\bibinfo {year} {1975})}\BibitemShut {NoStop}%
\bibitem [{\citenamefont {White}(1992)}]{white_density_1992}%
  \BibitemOpen
  \bibfield  {author} {\bibinfo {author} {\bibfnamefont {S.~R.}\ \bibnamefont
  {White}},\ }\bibfield  {title} {\bibinfo {title} {Density matrix formulation
  for quantum renormalization groups},\ }\href
  {https://doi.org/10.1103/PhysRevLett.69.2863} {\bibfield  {journal} {\bibinfo
   {journal} {Phys. Rev. Lett.}\ }\textbf {\bibinfo {volume} {69}},\ \bibinfo
  {pages} {2863} (\bibinfo {year} {1992})}\BibitemShut {NoStop}%
\bibitem [{\citenamefont {Maier}\ \emph {et~al.}(2000)\citenamefont {Maier},
  \citenamefont {Jarrell}, \citenamefont {Pruschke},\ and\ \citenamefont
  {Keller}}]{maier_d_2000}%
  \BibitemOpen
  \bibfield  {author} {\bibinfo {author} {\bibfnamefont {T.}~\bibnamefont
  {Maier}}, \bibinfo {author} {\bibfnamefont {M.}~\bibnamefont {Jarrell}},
  \bibinfo {author} {\bibfnamefont {T.}~\bibnamefont {Pruschke}},\ and\
  \bibinfo {author} {\bibfnamefont {J.}~\bibnamefont {Keller}},\ }\bibfield
  {title} {\bibinfo {title} {{$d$} -{Wave} {Superconductivity} in the {Hubbard}
  {Model}},\ }\href {https://doi.org/10.1103/PhysRevLett.85.1524} {\bibfield
  {journal} {\bibinfo  {journal} {Phys. Rev. Lett.}\ }\textbf {\bibinfo
  {volume} {85}},\ \bibinfo {pages} {1524} (\bibinfo {year}
  {2000})}\BibitemShut {NoStop}%
\bibitem [{\citenamefont {Becca}\ \emph {et~al.}(2001)\citenamefont {Becca},
  \citenamefont {Capriotti},\ and\ \citenamefont
  {Sorella}}]{becca_stripes_2001}%
  \BibitemOpen
  \bibfield  {author} {\bibinfo {author} {\bibfnamefont {F.}~\bibnamefont
  {Becca}}, \bibinfo {author} {\bibfnamefont {L.}~\bibnamefont {Capriotti}},\
  and\ \bibinfo {author} {\bibfnamefont {S.}~\bibnamefont {Sorella}},\
  }\bibfield  {title} {\bibinfo {title} {Stripes and {Spin}
  {Incommensurabilities} {Are} {Favored} by {Lattice} {Anisotropies}},\ }\href
  {https://doi.org/10.1103/PhysRevLett.87.167005} {\bibfield  {journal}
  {\bibinfo  {journal} {Phys. Rev. Lett.}\ }\textbf {\bibinfo {volume} {87}},\
  \bibinfo {pages} {167005} (\bibinfo {year} {2001})}\BibitemShut {NoStop}%
\bibitem [{\citenamefont {Sorella}\ \emph {et~al.}(2002)\citenamefont
  {Sorella}, \citenamefont {Martins}, \citenamefont {Becca}, \citenamefont
  {Gazza}, \citenamefont {Capriotti}, \citenamefont {Parola},\ and\
  \citenamefont {Dagotto}}]{sorella_superconductivity_2002}%
  \BibitemOpen
  \bibfield  {author} {\bibinfo {author} {\bibfnamefont {S.}~\bibnamefont
  {Sorella}}, \bibinfo {author} {\bibfnamefont {G.~B.}\ \bibnamefont
  {Martins}}, \bibinfo {author} {\bibfnamefont {F.}~\bibnamefont {Becca}},
  \bibinfo {author} {\bibfnamefont {C.}~\bibnamefont {Gazza}}, \bibinfo
  {author} {\bibfnamefont {L.}~\bibnamefont {Capriotti}}, \bibinfo {author}
  {\bibfnamefont {A.}~\bibnamefont {Parola}},\ and\ \bibinfo {author}
  {\bibfnamefont {E.}~\bibnamefont {Dagotto}},\ }\bibfield  {title} {\bibinfo
  {title} {Superconductivity in the two-dimensional {$t$}-{$J$} model},\ }\href
  {https://doi.org/10.1103/PhysRevLett.88.117002} {\bibfield  {journal}
  {\bibinfo  {journal} {Phys. Rev. Lett.}\ }\textbf {\bibinfo {volume} {88}},\
  \bibinfo {pages} {117002} (\bibinfo {year} {2002})}\BibitemShut {NoStop}%
\bibitem [{\citenamefont {Anisimov}\ \emph {et~al.}(2002)\citenamefont
  {Anisimov}, \citenamefont {Korotin}, \citenamefont {Nekrasov}, \citenamefont
  {Pchelkina},\ and\ \citenamefont {Sorella}}]{anisimov_first_2002}%
  \BibitemOpen
  \bibfield  {author} {\bibinfo {author} {\bibfnamefont {V.~I.}\ \bibnamefont
  {Anisimov}}, \bibinfo {author} {\bibfnamefont {M.~A.}\ \bibnamefont
  {Korotin}}, \bibinfo {author} {\bibfnamefont {I.~A.}\ \bibnamefont
  {Nekrasov}}, \bibinfo {author} {\bibfnamefont {Z.~V.}\ \bibnamefont
  {Pchelkina}},\ and\ \bibinfo {author} {\bibfnamefont {S.}~\bibnamefont
  {Sorella}},\ }\bibfield  {title} {\bibinfo {title} {First principles
  electronic model for high-temperature superconductivity},\ }\href
  {https://doi.org/10.1103/PhysRevB.66.100502} {\bibfield  {journal} {\bibinfo
  {journal} {Phys. Rev. B}\ }\textbf {\bibinfo {volume} {66}},\ \bibinfo
  {pages} {100502(R)} (\bibinfo {year} {2002})}\BibitemShut {NoStop}%
\bibitem [{\citenamefont {Spanu}\ \emph {et~al.}(2008)\citenamefont {Spanu},
  \citenamefont {Lugas}, \citenamefont {Becca},\ and\ \citenamefont
  {Sorella}}]{spanu_magnetism_2008}%
  \BibitemOpen
  \bibfield  {author} {\bibinfo {author} {\bibfnamefont {L.}~\bibnamefont
  {Spanu}}, \bibinfo {author} {\bibfnamefont {M.}~\bibnamefont {Lugas}},
  \bibinfo {author} {\bibfnamefont {F.}~\bibnamefont {Becca}},\ and\ \bibinfo
  {author} {\bibfnamefont {S.}~\bibnamefont {Sorella}},\ }\bibfield  {title}
  {\bibinfo {title} {Magnetism and superconductivity in the
  {$t$}-{$t^{\prime}$}-{$J$} model},\ }\href
  {https://doi.org/10.1103/PhysRevB.77.024510} {\bibfield  {journal} {\bibinfo
  {journal} {Phys. Rev. B}\ }\textbf {\bibinfo {volume} {77}},\ \bibinfo
  {pages} {024510} (\bibinfo {year} {2008})}\BibitemShut {NoStop}%
\bibitem [{\citenamefont {Ashkenazi}(2011)}]{ashkenazi_theory_2011}%
  \BibitemOpen
  \bibfield  {author} {\bibinfo {author} {\bibfnamefont {J.}~\bibnamefont
  {Ashkenazi}},\ }\bibfield  {title} {\bibinfo {title} {A {Theory} for the
  {High-$T_c$} {Cuprates}: {Anomalous} {Normal}-{State} and {Spectroscopic}
  {Properties}, {Phase} {Diagram}, and {Pairing}},\ }\href
  {https://doi.org/10.1007/s10948-010-0823-8} {\bibfield  {journal} {\bibinfo
  {journal} {J. Supercond. Nov. Magn.}\ }\textbf {\bibinfo {volume} {24}},\
  \bibinfo {pages} {1281} (\bibinfo {year} {2011})}\BibitemShut {NoStop}%
\bibitem [{\citenamefont {Hirayama}\ \emph {et~al.}(2013)\citenamefont
  {Hirayama}, \citenamefont {Miyake},\ and\ \citenamefont
  {Imada}}]{hirayama_derivation_2013}%
  \BibitemOpen
  \bibfield  {author} {\bibinfo {author} {\bibfnamefont {M.}~\bibnamefont
  {Hirayama}}, \bibinfo {author} {\bibfnamefont {T.}~\bibnamefont {Miyake}},\
  and\ \bibinfo {author} {\bibfnamefont {M.}~\bibnamefont {Imada}},\ }\bibfield
   {title} {\bibinfo {title} {Derivation of static low-energy effective models
  by an ab initio downfolding method without double counting of {Coulomb}
  correlations: {Application} to {SrVO}${}_{3}$, {FeSe}, and {FeTe}},\ }\href
  {https://doi.org/10.1103/PhysRevB.87.195144} {\bibfield  {journal} {\bibinfo
  {journal} {Phys. Rev. B}\ }\textbf {\bibinfo {volume} {87}},\ \bibinfo
  {pages} {195144} (\bibinfo {year} {2013})}\BibitemShut {NoStop}%
\bibitem [{\citenamefont {Qin}\ \emph {et~al.}(2020)\citenamefont {Qin},
  \citenamefont {Chung}, \citenamefont {Shi}, \citenamefont {Vitali},
  \citenamefont {Hubig}, \citenamefont {Schollw\"ock}, \citenamefont {White},\
  and\ \citenamefont {Zhang}}]{qin_absence_2020}%
  \BibitemOpen
  \bibfield  {author} {\bibinfo {author} {\bibfnamefont {M.}~\bibnamefont
  {Qin}}, \bibinfo {author} {\bibfnamefont {C.-M.}\ \bibnamefont {Chung}},
  \bibinfo {author} {\bibfnamefont {H.}~\bibnamefont {Shi}}, \bibinfo {author}
  {\bibfnamefont {E.}~\bibnamefont {Vitali}}, \bibinfo {author} {\bibfnamefont
  {C.}~\bibnamefont {Hubig}}, \bibinfo {author} {\bibfnamefont
  {U.}~\bibnamefont {Schollw\"ock}}, \bibinfo {author} {\bibfnamefont {S.~R.}\
  \bibnamefont {White}},\ and\ \bibinfo {author} {\bibfnamefont
  {S.}~\bibnamefont {Zhang}} (\bibinfo {collaboration} {Simons Collaboration on
  the Many-Electron Problem}),\ }\bibfield  {title} {\bibinfo {title} {Absence
  of {Superconductivity} in the {Pure} {Two}-{Dimensional} {Hubbard} {Model}},\
  }\href {https://doi.org/10.1103/PhysRevX.10.031016} {\bibfield  {journal}
  {\bibinfo  {journal} {Phys. Rev. X}\ }\textbf {\bibinfo {volume} {10}},\
  \bibinfo {pages} {031016} (\bibinfo {year} {2020})}\BibitemShut {NoStop}%
\bibitem [{\citenamefont {Jiang}\ \emph {et~al.}(2021)\citenamefont {Jiang},
  \citenamefont {Scalapino},\ and\ \citenamefont
  {White}}]{jiang_ground-state_2021}%
  \BibitemOpen
  \bibfield  {author} {\bibinfo {author} {\bibfnamefont {S.}~\bibnamefont
  {Jiang}}, \bibinfo {author} {\bibfnamefont {D.~J.}\ \bibnamefont
  {Scalapino}},\ and\ \bibinfo {author} {\bibfnamefont {S.~R.}\ \bibnamefont
  {White}},\ }\bibfield  {title} {\bibinfo {title} {Ground-state phase diagram
  of the {$t$-$t'$-$J$} model},\ }\href
  {https://doi.org/10.1073/pnas.2109978118} {\bibfield  {journal} {\bibinfo
  {journal} {Proc. Natl. Acad. Sci. U.S.A.}\ }\textbf {\bibinfo {volume}
  {118}},\ \bibinfo {pages} {e2109978118} (\bibinfo {year} {2021})}\BibitemShut
  {NoStop}%
\bibitem [{\citenamefont {Jiang}\ \emph {et~al.}(2022)\citenamefont {Jiang},
  \citenamefont {Scalapino},\ and\ \citenamefont {White}}]{jiang_pairing_2022}%
  \BibitemOpen
  \bibfield  {author} {\bibinfo {author} {\bibfnamefont {S.}~\bibnamefont
  {Jiang}}, \bibinfo {author} {\bibfnamefont {D.~J.}\ \bibnamefont
  {Scalapino}},\ and\ \bibinfo {author} {\bibfnamefont {S.~R.}\ \bibnamefont
  {White}},\ }\bibfield  {title} {\bibinfo {title} {Pairing properties of the
  {$t$-$t'$-$t''$-$J$} model},\ }\href
  {https://doi.org/10.1103/PhysRevB.106.174507} {\bibfield  {journal} {\bibinfo
   {journal} {Phys. Rev. B}\ }\textbf {\bibinfo {volume} {106}},\ \bibinfo
  {pages} {174507} (\bibinfo {year} {2022})}\BibitemShut {NoStop}%
\bibitem [{\citenamefont {Witczak-Krempa}\ \emph {et~al.}(2014)\citenamefont
  {Witczak-Krempa}, \citenamefont {Chen}, \citenamefont {Kim},\ and\
  \citenamefont {Balents}}]{witczak-krempa_correlated_2014}%
  \BibitemOpen
  \bibfield  {author} {\bibinfo {author} {\bibfnamefont {W.}~\bibnamefont
  {Witczak-Krempa}}, \bibinfo {author} {\bibfnamefont {G.}~\bibnamefont
  {Chen}}, \bibinfo {author} {\bibfnamefont {Y.~B.}\ \bibnamefont {Kim}},\ and\
  \bibinfo {author} {\bibfnamefont {L.}~\bibnamefont {Balents}},\ }\bibfield
  {title} {\bibinfo {title} {Correlated {Quantum} {Phenomena} in the {Strong}
  {Spin}-{Orbit} {Regime}},\ }\href
  {https://doi.org/10.1146/annurev-conmatphys-020911-125138} {\bibfield
  {journal} {\bibinfo  {journal} {Annu. Rev. Condens. Matter Phys.}\ }\textbf
  {\bibinfo {volume} {5}},\ \bibinfo {pages} {57} (\bibinfo {year}
  {2014})}\BibitemShut {NoStop}%
\bibitem [{\citenamefont {Rau}\ \emph {et~al.}(2016)\citenamefont {Rau},
  \citenamefont {Lee},\ and\ \citenamefont {Kee}}]{rau_spin-orbit_2016}%
  \BibitemOpen
  \bibfield  {author} {\bibinfo {author} {\bibfnamefont {J.~G.}\ \bibnamefont
  {Rau}}, \bibinfo {author} {\bibfnamefont {E.~K.-H.}\ \bibnamefont {Lee}},\
  and\ \bibinfo {author} {\bibfnamefont {H.-Y.}\ \bibnamefont {Kee}},\
  }\bibfield  {title} {\bibinfo {title} {Spin-{Orbit} {Physics} {Giving} {Rise}
  to {Novel} {Phases} in {Correlated} {Systems}: {Iridates} and {Related}
  {Materials}},\ }\href
  {https://doi.org/10.1146/annurev-conmatphys-031115-011319} {\bibfield
  {journal} {\bibinfo  {journal} {Annu. Rev. Condens. Matter Phys.}\ }\textbf
  {\bibinfo {volume} {7}},\ \bibinfo {pages} {195} (\bibinfo {year}
  {2016})}\BibitemShut {NoStop}%
\bibitem [{\citenamefont {Banerjee}\ \emph {et~al.}(2017)\citenamefont
  {Banerjee}, \citenamefont {Yan}, \citenamefont {Knolle}, \citenamefont
  {Bridges}, \citenamefont {Stone}, \citenamefont {Lumsden}, \citenamefont
  {Mandrus}, \citenamefont {Tennant}, \citenamefont {Moessner},\ and\
  \citenamefont {Nagler}}]{banerjee_neutron_2017}%
  \BibitemOpen
  \bibfield  {author} {\bibinfo {author} {\bibfnamefont {A.}~\bibnamefont
  {Banerjee}}, \bibinfo {author} {\bibfnamefont {J.}~\bibnamefont {Yan}},
  \bibinfo {author} {\bibfnamefont {J.}~\bibnamefont {Knolle}}, \bibinfo
  {author} {\bibfnamefont {C.~A.}\ \bibnamefont {Bridges}}, \bibinfo {author}
  {\bibfnamefont {M.~B.}\ \bibnamefont {Stone}}, \bibinfo {author}
  {\bibfnamefont {M.~D.}\ \bibnamefont {Lumsden}}, \bibinfo {author}
  {\bibfnamefont {D.~G.}\ \bibnamefont {Mandrus}}, \bibinfo {author}
  {\bibfnamefont {D.~A.}\ \bibnamefont {Tennant}}, \bibinfo {author}
  {\bibfnamefont {R.}~\bibnamefont {Moessner}},\ and\ \bibinfo {author}
  {\bibfnamefont {S.~E.}\ \bibnamefont {Nagler}},\ }\bibfield  {title}
  {\bibinfo {title} {Neutron scattering in the proximate quantum spin liquid
  $\alpha$-{RuCl}$_{\textrm{3}}$},\ }\href
  {https://doi.org/10.1126/science.aah6015} {\bibfield  {journal} {\bibinfo
  {journal} {Science}\ }\textbf {\bibinfo {volume} {356}},\ \bibinfo {pages}
  {1055} (\bibinfo {year} {2017})}\BibitemShut {NoStop}%
\bibitem [{\citenamefont {Savary}\ and\ \citenamefont
  {Balents}(2017)}]{savary_quantum_2017}%
  \BibitemOpen
  \bibfield  {author} {\bibinfo {author} {\bibfnamefont {L.}~\bibnamefont
  {Savary}}\ and\ \bibinfo {author} {\bibfnamefont {L.}~\bibnamefont
  {Balents}},\ }\bibfield  {title} {\bibinfo {title} {Quantum spin liquids: a
  review},\ }\href {https://doi.org/10.1088/0034-4885/80/1/016502} {\bibfield
  {journal} {\bibinfo  {journal} {Rep. Prog. Phys.}\ }\textbf {\bibinfo
  {volume} {80}},\ \bibinfo {pages} {016502} (\bibinfo {year}
  {2017})}\BibitemShut {NoStop}%
\bibitem [{\citenamefont {Zhou}\ \emph {et~al.}(2017)\citenamefont {Zhou},
  \citenamefont {Kanoda},\ and\ \citenamefont {Ng}}]{zhou_quantum_2017}%
  \BibitemOpen
  \bibfield  {author} {\bibinfo {author} {\bibfnamefont {Y.}~\bibnamefont
  {Zhou}}, \bibinfo {author} {\bibfnamefont {K.}~\bibnamefont {Kanoda}},\ and\
  \bibinfo {author} {\bibfnamefont {T.-K.}\ \bibnamefont {Ng}},\ }\bibfield
  {title} {\bibinfo {title} {Quantum spin liquid states},\ }\href
  {https://doi.org/10.1103/RevModPhys.89.025003} {\bibfield  {journal}
  {\bibinfo  {journal} {Rev. Mod. Phys.}\ }\textbf {\bibinfo {volume} {89}},\
  \bibinfo {pages} {025003} (\bibinfo {year} {2017})}\BibitemShut {NoStop}%
\bibitem [{\citenamefont {Andrei}\ and\ \citenamefont
  {MacDonald}(2020)}]{andrei_graphene_2020}%
  \BibitemOpen
  \bibfield  {author} {\bibinfo {author} {\bibfnamefont {E.~Y.}\ \bibnamefont
  {Andrei}}\ and\ \bibinfo {author} {\bibfnamefont {A.~H.}\ \bibnamefont
  {MacDonald}},\ }\bibfield  {title} {\bibinfo {title} {Graphene bilayers with
  a twist},\ }\href {https://doi.org/10.1038/s41563-020-00840-0} {\bibfield
  {journal} {\bibinfo  {journal} {Nat. Mat.}\ }\textbf {\bibinfo {volume}
  {19}},\ \bibinfo {pages} {1265} (\bibinfo {year} {2020})}\BibitemShut
  {NoStop}%
\bibitem [{\citenamefont {Andrei}\ \emph {et~al.}(2021)\citenamefont {Andrei},
  \citenamefont {Efetov}, \citenamefont {Jarillo-Herrero}, \citenamefont
  {MacDonald}, \citenamefont {Mak}, \citenamefont {Senthil}, \citenamefont
  {Tutuc}, \citenamefont {Yazdani},\ and\ \citenamefont
  {Young}}]{andrei_marvels_2021}%
  \BibitemOpen
  \bibfield  {author} {\bibinfo {author} {\bibfnamefont {E.~Y.}\ \bibnamefont
  {Andrei}}, \bibinfo {author} {\bibfnamefont {D.~K.}\ \bibnamefont {Efetov}},
  \bibinfo {author} {\bibfnamefont {P.}~\bibnamefont {Jarillo-Herrero}},
  \bibinfo {author} {\bibfnamefont {A.~H.}\ \bibnamefont {MacDonald}}, \bibinfo
  {author} {\bibfnamefont {K.~F.}\ \bibnamefont {Mak}}, \bibinfo {author}
  {\bibfnamefont {T.}~\bibnamefont {Senthil}}, \bibinfo {author} {\bibfnamefont
  {E.}~\bibnamefont {Tutuc}}, \bibinfo {author} {\bibfnamefont
  {A.}~\bibnamefont {Yazdani}},\ and\ \bibinfo {author} {\bibfnamefont {A.~F.}\
  \bibnamefont {Young}},\ }\bibfield  {title} {\bibinfo {title} {The marvels of
  moir{\'e} materials},\ }\href {https://doi.org/10.1038/s41578-021-00284-1}
  {\bibfield  {journal} {\bibinfo  {journal} {Nat. Rev. Mater.}\ }\textbf
  {\bibinfo {volume} {6}},\ \bibinfo {pages} {201} (\bibinfo {year}
  {2021})}\BibitemShut {NoStop}%
\bibitem [{\citenamefont {Aryasetiawan}\ \emph {et~al.}(2004)\citenamefont
  {Aryasetiawan}, \citenamefont {Imada}, \citenamefont {Georges}, \citenamefont
  {Kotliar}, \citenamefont {Biermann},\ and\ \citenamefont
  {Lichtenstein}}]{aryasetiawan_frequency-dependent_2004}%
  \BibitemOpen
  \bibfield  {author} {\bibinfo {author} {\bibfnamefont {F.}~\bibnamefont
  {Aryasetiawan}}, \bibinfo {author} {\bibfnamefont {M.}~\bibnamefont {Imada}},
  \bibinfo {author} {\bibfnamefont {A.}~\bibnamefont {Georges}}, \bibinfo
  {author} {\bibfnamefont {G.}~\bibnamefont {Kotliar}}, \bibinfo {author}
  {\bibfnamefont {S.}~\bibnamefont {Biermann}},\ and\ \bibinfo {author}
  {\bibfnamefont {A.~I.}\ \bibnamefont {Lichtenstein}},\ }\bibfield  {title}
  {\bibinfo {title} {Frequency-dependent local interactions and low-energy
  effective models from electronic structure calculations},\ }\href
  {https://doi.org/10.1103/PhysRevB.70.195104} {\bibfield  {journal} {\bibinfo
  {journal} {Phys. Rev. B}\ }\textbf {\bibinfo {volume} {70}},\ \bibinfo
  {pages} {195104} (\bibinfo {year} {2004})}\BibitemShut {NoStop}%
\bibitem [{\citenamefont {Aryasetiawan}\ \emph {et~al.}(2006)\citenamefont
  {Aryasetiawan}, \citenamefont {Karlsson}, \citenamefont {Jepsen},\ and\
  \citenamefont {Sch{\"o}nberger}}]{aryasetiawan_calculations_2006}%
  \BibitemOpen
  \bibfield  {author} {\bibinfo {author} {\bibfnamefont {F.}~\bibnamefont
  {Aryasetiawan}}, \bibinfo {author} {\bibfnamefont {K.}~\bibnamefont
  {Karlsson}}, \bibinfo {author} {\bibfnamefont {O.}~\bibnamefont {Jepsen}},\
  and\ \bibinfo {author} {\bibfnamefont {U.}~\bibnamefont {Sch{\"o}nberger}},\
  }\bibfield  {title} {\bibinfo {title} {Calculations of {Hubbard} {$U$} from
  first-principles},\ }\href {https://doi.org/10.1103/PhysRevB.74.125106}
  {\bibfield  {journal} {\bibinfo  {journal} {Phys. Rev. B}\ }\textbf {\bibinfo
  {volume} {74}},\ \bibinfo {pages} {125106} (\bibinfo {year}
  {2006})}\BibitemShut {NoStop}%
\bibitem [{\citenamefont {Scott}\ and\ \citenamefont
  {Booth}(2024)}]{scott_rigorous_2024}%
  \BibitemOpen
  \bibfield  {author} {\bibinfo {author} {\bibfnamefont {C.~J.~C.}\
  \bibnamefont {Scott}}\ and\ \bibinfo {author} {\bibfnamefont {G.~H.}\
  \bibnamefont {Booth}},\ }\bibfield  {title} {\bibinfo {title} {Rigorous
  screened interactions for realistic correlated electron systems},\ }\href
  {https://doi.org/10.1103/PhysRevLett.132.076401} {\bibfield  {journal}
  {\bibinfo  {journal} {Phys. Rev. Lett.}\ }\textbf {\bibinfo {volume} {132}},\
  \bibinfo {pages} {076401} (\bibinfo {year} {2024})}\BibitemShut {NoStop}%
\bibitem [{\citenamefont {Hedin}(1965)}]{hedin_new_1965}%
  \BibitemOpen
  \bibfield  {author} {\bibinfo {author} {\bibfnamefont {L.}~\bibnamefont
  {Hedin}},\ }\bibfield  {title} {\bibinfo {title} {New {Method} for
  {Calculating} the {One}-{Particle} {Green}'s {Function} with {Application} to
  the {Electron}-{Gas} {Problem}},\ }\href
  {https://doi.org/10.1103/PhysRev.139.A796} {\bibfield  {journal} {\bibinfo
  {journal} {Phys. Rev.}\ }\textbf {\bibinfo {volume} {139}},\ \bibinfo {pages}
  {A796} (\bibinfo {year} {1965})}\BibitemShut {NoStop}%
\bibitem [{\citenamefont {Aryasetiawan}\ and\ \citenamefont
  {Gunnarsson}(1998)}]{aryasetiawan_gw_1998}%
  \BibitemOpen
  \bibfield  {author} {\bibinfo {author} {\bibfnamefont {F.}~\bibnamefont
  {Aryasetiawan}}\ and\ \bibinfo {author} {\bibfnamefont {O.}~\bibnamefont
  {Gunnarsson}},\ }\bibfield  {title} {\bibinfo {title} {The \textbf{
  \textit{{GW}} } method},\ }\href {https://doi.org/10.1088/0034-4885/61/3/002}
  {\bibfield  {journal} {\bibinfo  {journal} {Rep. Prog. Phys.}\ }\textbf
  {\bibinfo {volume} {61}},\ \bibinfo {pages} {237} (\bibinfo {year}
  {1998})}\BibitemShut {NoStop}%
\bibitem [{\citenamefont {Romanova}\ \emph {et~al.}(2023)\citenamefont
  {Romanova}, \citenamefont {Weng}, \citenamefont {Apelian},\ and\
  \citenamefont {Vlček}}]{romanova_dynamical_2023}%
  \BibitemOpen
  \bibfield  {author} {\bibinfo {author} {\bibfnamefont {M.}~\bibnamefont
  {Romanova}}, \bibinfo {author} {\bibfnamefont {G.}~\bibnamefont {Weng}},
  \bibinfo {author} {\bibfnamefont {A.}~\bibnamefont {Apelian}},\ and\ \bibinfo
  {author} {\bibfnamefont {V.}~\bibnamefont {Vlček}},\ }\bibfield  {title}
  {\bibinfo {title} {Dynamical downfolding for localized quantum states},\
  }\href {https://doi.org/10.1038/s41524-023-01078-5} {\bibfield  {journal}
  {\bibinfo  {journal} {npj Comput. Mater.}\ }\textbf {\bibinfo {volume} {9}},\
  \bibinfo {pages} {126} (\bibinfo {year} {2023})}\BibitemShut {NoStop}%
\bibitem [{\citenamefont {Canestraight}\ \emph {et~al.}(2024)\citenamefont
  {Canestraight}, \citenamefont {Lei}, \citenamefont {Ibrahim},\ and\
  \citenamefont {Vlček}}]{canestraight_efficient_2024}%
  \BibitemOpen
  \bibfield  {author} {\bibinfo {author} {\bibfnamefont {A.}~\bibnamefont
  {Canestraight}}, \bibinfo {author} {\bibfnamefont {X.}~\bibnamefont {Lei}},
  \bibinfo {author} {\bibfnamefont {K.~Z.}\ \bibnamefont {Ibrahim}},\ and\
  \bibinfo {author} {\bibfnamefont {V.}~\bibnamefont {Vlček}},\ }\bibfield
  {title} {\bibinfo {title} {Efficient {Quasiparticle} {Determination} beyond
  the {Diagonal} {Approximation} via {Random} {Compression}},\ }\href
  {https://doi.org/10.1021/acs.jctc.3c01069} {\bibfield  {journal} {\bibinfo
  {journal} {J. Chem. Theory Comput.}\ }\textbf {\bibinfo {volume} {20}},\
  \bibinfo {pages} {551} (\bibinfo {year} {2024})}\BibitemShut {NoStop}%
\bibitem [{\citenamefont {Cohen}\ \emph {et~al.}(2008)\citenamefont {Cohen},
  \citenamefont {Mori-Sánchez},\ and\ \citenamefont
  {Yang}}]{cohen_insights_2008}%
  \BibitemOpen
  \bibfield  {author} {\bibinfo {author} {\bibfnamefont {A.~J.}\ \bibnamefont
  {Cohen}}, \bibinfo {author} {\bibfnamefont {P.}~\bibnamefont
  {Mori-Sánchez}},\ and\ \bibinfo {author} {\bibfnamefont {W.}~\bibnamefont
  {Yang}},\ }\bibfield  {title} {\bibinfo {title} {Insights into {Current}
  {Limitations} of {Density} {Functional} {Theory}},\ }\href
  {https://doi.org/10.1126/science.1158722} {\bibfield  {journal} {\bibinfo
  {journal} {Science}\ }\textbf {\bibinfo {volume} {321}},\ \bibinfo {pages}
  {792} (\bibinfo {year} {2008})}\BibitemShut {NoStop}%
\bibitem [{\citenamefont {Williams}\ \emph {et~al.}(2020)\citenamefont
  {Williams}, \citenamefont {Yao}, \citenamefont {Li}, \citenamefont {Chen},
  \citenamefont {Shi}, \citenamefont {Motta}, \citenamefont {Niu},
  \citenamefont {Ray}, \citenamefont {Guo}, \citenamefont {Anderson},
  \citenamefont {Li}, \citenamefont {Tran}, \citenamefont {Yeh}, \citenamefont
  {Mussard}, \citenamefont {Sharma}, \citenamefont {Bruneval}, \citenamefont
  {van Schilfgaarde}, \citenamefont {Booth}, \citenamefont {Chan},
  \citenamefont {Zhang}, \citenamefont {Gull}, \citenamefont {Zgid},
  \citenamefont {Millis}, \citenamefont {Umrigar},\ and\ \citenamefont
  {Wagner}}]{williams_direct_2020}%
  \BibitemOpen
  \bibfield  {author} {\bibinfo {author} {\bibfnamefont {K.~T.}\ \bibnamefont
  {Williams}}, \bibinfo {author} {\bibfnamefont {Y.}~\bibnamefont {Yao}},
  \bibinfo {author} {\bibfnamefont {J.}~\bibnamefont {Li}}, \bibinfo {author}
  {\bibfnamefont {L.}~\bibnamefont {Chen}}, \bibinfo {author} {\bibfnamefont
  {H.}~\bibnamefont {Shi}}, \bibinfo {author} {\bibfnamefont {M.}~\bibnamefont
  {Motta}}, \bibinfo {author} {\bibfnamefont {C.}~\bibnamefont {Niu}}, \bibinfo
  {author} {\bibfnamefont {U.}~\bibnamefont {Ray}}, \bibinfo {author}
  {\bibfnamefont {S.}~\bibnamefont {Guo}}, \bibinfo {author} {\bibfnamefont
  {R.~J.}\ \bibnamefont {Anderson}}, \bibinfo {author} {\bibfnamefont
  {J.}~\bibnamefont {Li}}, \bibinfo {author} {\bibfnamefont {L.~N.}\
  \bibnamefont {Tran}}, \bibinfo {author} {\bibfnamefont {C.-N.}\ \bibnamefont
  {Yeh}}, \bibinfo {author} {\bibfnamefont {B.}~\bibnamefont {Mussard}},
  \bibinfo {author} {\bibfnamefont {S.}~\bibnamefont {Sharma}}, \bibinfo
  {author} {\bibfnamefont {F.}~\bibnamefont {Bruneval}}, \bibinfo {author}
  {\bibfnamefont {M.}~\bibnamefont {van Schilfgaarde}}, \bibinfo {author}
  {\bibfnamefont {G.~H.}\ \bibnamefont {Booth}}, \bibinfo {author}
  {\bibfnamefont {G.~K.-L.}\ \bibnamefont {Chan}}, \bibinfo {author}
  {\bibfnamefont {S.}~\bibnamefont {Zhang}}, \bibinfo {author} {\bibfnamefont
  {E.}~\bibnamefont {Gull}}, \bibinfo {author} {\bibfnamefont {D.}~\bibnamefont
  {Zgid}}, \bibinfo {author} {\bibfnamefont {A.}~\bibnamefont {Millis}},
  \bibinfo {author} {\bibfnamefont {C.~J.}\ \bibnamefont {Umrigar}},\ and\
  \bibinfo {author} {\bibfnamefont {L.~K.}\ \bibnamefont {Wagner}} (\bibinfo
  {collaboration} {Simons Collaboration on the Many-Electron Problem}),\
  }\bibfield  {title} {\bibinfo {title} {Direct {Comparison} of {Many}-{Body}
  {Methods} for {Realistic} {Electronic} {Hamiltonians}},\ }\href
  {https://doi.org/10.1103/PhysRevX.10.011041} {\bibfield  {journal} {\bibinfo
  {journal} {Phys. Rev. X}\ }\textbf {\bibinfo {volume} {10}},\ \bibinfo
  {pages} {011041} (\bibinfo {year} {2020})}\BibitemShut {NoStop}%
\bibitem [{\citenamefont {Chang}\ \emph {et~al.}(2024)\citenamefont {Chang},
  \citenamefont {Van~Loon}, \citenamefont {Eskridge}, \citenamefont
  {Busemeyer}, \citenamefont {Morales}, \citenamefont {Dreyer}, \citenamefont
  {Millis}, \citenamefont {Zhang}, \citenamefont {Wehling}, \citenamefont
  {Wagner},\ and\ \citenamefont {R{\"o}sner}}]{chang_downfolding_2024}%
  \BibitemOpen
  \bibfield  {author} {\bibinfo {author} {\bibfnamefont {Y.}~\bibnamefont
  {Chang}}, \bibinfo {author} {\bibfnamefont {E.~G. C.~P.}\ \bibnamefont
  {Van~Loon}}, \bibinfo {author} {\bibfnamefont {B.}~\bibnamefont {Eskridge}},
  \bibinfo {author} {\bibfnamefont {B.}~\bibnamefont {Busemeyer}}, \bibinfo
  {author} {\bibfnamefont {M.~A.}\ \bibnamefont {Morales}}, \bibinfo {author}
  {\bibfnamefont {C.~E.}\ \bibnamefont {Dreyer}}, \bibinfo {author}
  {\bibfnamefont {A.~J.}\ \bibnamefont {Millis}}, \bibinfo {author}
  {\bibfnamefont {S.}~\bibnamefont {Zhang}}, \bibinfo {author} {\bibfnamefont
  {T.~O.}\ \bibnamefont {Wehling}}, \bibinfo {author} {\bibfnamefont {L.~K.}\
  \bibnamefont {Wagner}},\ and\ \bibinfo {author} {\bibfnamefont
  {M.}~\bibnamefont {R{\"o}sner}},\ }\bibfield  {title} {\bibinfo {title}
  {Downfolding from ab initio to interacting model {Hamiltonians}:
  comprehensive analysis and benchmarking of the {DFT}+{cRPA} approach},\
  }\href {https://doi.org/10.1038/s41524-024-01314-6} {\bibfield  {journal}
  {\bibinfo  {journal} {npj Computational Materials}\ }\textbf {\bibinfo
  {volume} {10}},\ \bibinfo {pages} {129} (\bibinfo {year} {2024})}\BibitemShut
  {NoStop}%
\bibitem [{\citenamefont {Zhang}\ \emph {et~al.}(2017)\citenamefont {Zhang},
  \citenamefont {Haule},\ and\ \citenamefont {Vanderbilt}}]{zhang_metal_2017}%
  \BibitemOpen
  \bibfield  {author} {\bibinfo {author} {\bibfnamefont {H.}~\bibnamefont
  {Zhang}}, \bibinfo {author} {\bibfnamefont {K.}~\bibnamefont {Haule}},\ and\
  \bibinfo {author} {\bibfnamefont {D.}~\bibnamefont {Vanderbilt}},\ }\bibfield
   {title} {\bibinfo {title} {Metal-{Insulator} {Transition} and {Topological}
  {Properties} of {Pyrochlore} {Iridates}},\ }\href
  {https://doi.org/10.1103/PhysRevLett.118.026404} {\bibfield  {journal}
  {\bibinfo  {journal} {Phys. Rev. Lett.}\ }\textbf {\bibinfo {volume} {118}},\
  \bibinfo {pages} {026404} (\bibinfo {year} {2017})}\BibitemShut {NoStop}%
\bibitem [{\citenamefont {Wang}\ \emph
  {et~al.}(2017{\natexlab{a}})\citenamefont {Wang}, \citenamefont {Go},\ and\
  \citenamefont {Millis}}]{wang_electron_2017}%
  \BibitemOpen
  \bibfield  {author} {\bibinfo {author} {\bibfnamefont {R.}~\bibnamefont
  {Wang}}, \bibinfo {author} {\bibfnamefont {A.}~\bibnamefont {Go}},\ and\
  \bibinfo {author} {\bibfnamefont {A.~J.}\ \bibnamefont {Millis}},\ }\bibfield
   {title} {\bibinfo {title} {Electron interactions, spin-orbit coupling, and
  intersite correlations in pyrochlore iridates},\ }\href
  {https://doi.org/10.1103/PhysRevB.95.045133} {\bibfield  {journal} {\bibinfo
  {journal} {Phys. Rev. B}\ }\textbf {\bibinfo {volume} {95}},\ \bibinfo
  {pages} {045133} (\bibinfo {year} {2017}{\natexlab{a}})}\BibitemShut
  {NoStop}%
\bibitem [{\citenamefont {Wang}\ \emph
  {et~al.}(2017{\natexlab{b}})\citenamefont {Wang}, \citenamefont {Go},\ and\
  \citenamefont {Millis}}]{wang_weyl_2017}%
  \BibitemOpen
  \bibfield  {author} {\bibinfo {author} {\bibfnamefont {R.}~\bibnamefont
  {Wang}}, \bibinfo {author} {\bibfnamefont {A.}~\bibnamefont {Go}},\ and\
  \bibinfo {author} {\bibfnamefont {A.}~\bibnamefont {Millis}},\ }\bibfield
  {title} {\bibinfo {title} {Weyl rings and enhanced susceptibilities in
  pyrochlore iridates: $k\ifmmode\cdot\else\textperiodcentered\fi{}p$ analysis
  of cluster dynamical mean-field theory results},\ }\href
  {https://doi.org/10.1103/PhysRevB.96.195158} {\bibfield  {journal} {\bibinfo
  {journal} {Phys. Rev. B}\ }\textbf {\bibinfo {volume} {96}},\ \bibinfo
  {pages} {195158} (\bibinfo {year} {2017}{\natexlab{b}})}\BibitemShut
  {NoStop}%
\bibitem [{\citenamefont {Liu}\ \emph {et~al.}(2021)\citenamefont {Liu},
  \citenamefont {Fang}, \citenamefont {Fu}, \citenamefont {Ge}, \citenamefont
  {Kareev}, \citenamefont {Kim}, \citenamefont {Choi}, \citenamefont
  {Karapetrova}, \citenamefont {Zhang}, \citenamefont {Gu}, \citenamefont
  {Choi}, \citenamefont {Wen}, \citenamefont {Wilson}, \citenamefont {Fabbris},
  \citenamefont {Ryan}, \citenamefont {Freeland}, \citenamefont {Haskel},
  \citenamefont {Wu}, \citenamefont {Pixley},\ and\ \citenamefont
  {Chakhalian}}]{liu_magnetic_2021}%
  \BibitemOpen
  \bibfield  {author} {\bibinfo {author} {\bibfnamefont {X.}~\bibnamefont
  {Liu}}, \bibinfo {author} {\bibfnamefont {S.}~\bibnamefont {Fang}}, \bibinfo
  {author} {\bibfnamefont {Y.}~\bibnamefont {Fu}}, \bibinfo {author}
  {\bibfnamefont {W.}~\bibnamefont {Ge}}, \bibinfo {author} {\bibfnamefont
  {M.}~\bibnamefont {Kareev}}, \bibinfo {author} {\bibfnamefont {J.-W.}\
  \bibnamefont {Kim}}, \bibinfo {author} {\bibfnamefont {Y.}~\bibnamefont
  {Choi}}, \bibinfo {author} {\bibfnamefont {E.}~\bibnamefont {Karapetrova}},
  \bibinfo {author} {\bibfnamefont {Q.}~\bibnamefont {Zhang}}, \bibinfo
  {author} {\bibfnamefont {L.}~\bibnamefont {Gu}}, \bibinfo {author}
  {\bibfnamefont {E.-S.}\ \bibnamefont {Choi}}, \bibinfo {author}
  {\bibfnamefont {F.}~\bibnamefont {Wen}}, \bibinfo {author} {\bibfnamefont
  {J.~H.}\ \bibnamefont {Wilson}}, \bibinfo {author} {\bibfnamefont
  {G.}~\bibnamefont {Fabbris}}, \bibinfo {author} {\bibfnamefont {P.~J.}\
  \bibnamefont {Ryan}}, \bibinfo {author} {\bibfnamefont {J.~W.}\ \bibnamefont
  {Freeland}}, \bibinfo {author} {\bibfnamefont {D.}~\bibnamefont {Haskel}},
  \bibinfo {author} {\bibfnamefont {W.}~\bibnamefont {Wu}}, \bibinfo {author}
  {\bibfnamefont {J.~H.}\ \bibnamefont {Pixley}},\ and\ \bibinfo {author}
  {\bibfnamefont {J.}~\bibnamefont {Chakhalian}},\ }\bibfield  {title}
  {\bibinfo {title} {Magnetic {Weyl} semimetallic phase in thin films of
  {Eu}$_2${Ir}$_2${O}$_7$},\ }\href
  {https://doi.org/10.1103/PhysRevLett.127.277204} {\bibfield  {journal}
  {\bibinfo  {journal} {Phys. Rev. Lett.}\ }\textbf {\bibinfo {volume} {127}},\
  \bibinfo {pages} {277204} (\bibinfo {year} {2021})}\BibitemShut {NoStop}%
\bibitem [{\citenamefont {Muechler}\ \emph {et~al.}(2022)\citenamefont
  {Muechler}, \citenamefont {Badrtdinov}, \citenamefont {Hampel}, \citenamefont
  {Cano}, \citenamefont {Rösner},\ and\ \citenamefont
  {Dreyer}}]{muechler_quantum_2022}%
  \BibitemOpen
  \bibfield  {author} {\bibinfo {author} {\bibfnamefont {L.}~\bibnamefont
  {Muechler}}, \bibinfo {author} {\bibfnamefont {D.~I.}\ \bibnamefont
  {Badrtdinov}}, \bibinfo {author} {\bibfnamefont {A.}~\bibnamefont {Hampel}},
  \bibinfo {author} {\bibfnamefont {J.}~\bibnamefont {Cano}}, \bibinfo {author}
  {\bibfnamefont {M.}~\bibnamefont {Rösner}},\ and\ \bibinfo {author}
  {\bibfnamefont {C.~E.}\ \bibnamefont {Dreyer}},\ }\bibfield  {title}
  {\bibinfo {title} {Quantum embedding methods for correlated excited states of
  point defects: {Case} studies and challenges},\ }\href
  {https://doi.org/10.1103/PhysRevB.105.235104} {\bibfield  {journal} {\bibinfo
   {journal} {Phys. Rev. B}\ }\textbf {\bibinfo {volume} {105}},\ \bibinfo
  {pages} {235104} (\bibinfo {year} {2022})}\BibitemShut {NoStop}%
\bibitem [{\citenamefont {Huron}\ \emph {et~al.}(1973)\citenamefont {Huron},
  \citenamefont {Malrieu},\ and\ \citenamefont
  {Rancurel}}]{huron_iterative_1973}%
  \BibitemOpen
  \bibfield  {author} {\bibinfo {author} {\bibfnamefont {B.}~\bibnamefont
  {Huron}}, \bibinfo {author} {\bibfnamefont {J.~P.}\ \bibnamefont {Malrieu}},\
  and\ \bibinfo {author} {\bibfnamefont {P.}~\bibnamefont {Rancurel}},\
  }\bibfield  {title} {\bibinfo {title} {Iterative perturbation calculations of
  ground and excited state energies from multiconfigurational zeroth‐order
  wavefunctions},\ }\href {https://doi.org/10.1063/1.1679199} {\bibfield
  {journal} {\bibinfo  {journal} {J. Chem. Phys.}\ }\textbf {\bibinfo {volume}
  {58}},\ \bibinfo {pages} {5745} (\bibinfo {year} {1973})}\BibitemShut
  {NoStop}%
\bibitem [{\citenamefont {Ruedenberg}\ \emph {et~al.}(1982)\citenamefont
  {Ruedenberg}, \citenamefont {Schmidt}, \citenamefont {Gilbert},\ and\
  \citenamefont {Elbert}}]{ruedenberg_are_1982}%
  \BibitemOpen
  \bibfield  {author} {\bibinfo {author} {\bibfnamefont {K.}~\bibnamefont
  {Ruedenberg}}, \bibinfo {author} {\bibfnamefont {M.~W.}\ \bibnamefont
  {Schmidt}}, \bibinfo {author} {\bibfnamefont {M.~M.}\ \bibnamefont
  {Gilbert}},\ and\ \bibinfo {author} {\bibfnamefont {S.}~\bibnamefont
  {Elbert}},\ }\bibfield  {title} {\bibinfo {title} {{Are} atoms intrinsic to
  molecular electronic wavefunctions? {I}. {T}he {FORS} model},\ }\href
  {https://doi.org/https://doi.org/10.1016/0301-0104(82)87004-3} {\bibfield
  {journal} {\bibinfo  {journal} {Chem. Phys.}\ }\textbf {\bibinfo {volume}
  {71}},\ \bibinfo {pages} {41} (\bibinfo {year} {1982})}\BibitemShut {NoStop}%
\bibitem [{\citenamefont {Cimiraglia}\ and\ \citenamefont
  {Persico}(1987)}]{cimiraglia_recent_1987}%
  \BibitemOpen
  \bibfield  {author} {\bibinfo {author} {\bibfnamefont {R.}~\bibnamefont
  {Cimiraglia}}\ and\ \bibinfo {author} {\bibfnamefont {M.}~\bibnamefont
  {Persico}},\ }\bibfield  {title} {\bibinfo {title} {Recent advances in
  multireference second order perturbation {CI}: {The} {CIPSI} method
  revisited},\ }\href {https://doi.org/10.1002/jcc.540080105} {\bibfield
  {journal} {\bibinfo  {journal} {J. Chem. Phys.}\ }\textbf {\bibinfo {volume}
  {8}},\ \bibinfo {pages} {39} (\bibinfo {year} {1987})}\BibitemShut {NoStop}%
\bibitem [{\citenamefont {Holmes}\ \emph {et~al.}(2016)\citenamefont {Holmes},
  \citenamefont {Tubman},\ and\ \citenamefont
  {Umrigar}}]{holmes_heat-bath_2016}%
  \BibitemOpen
  \bibfield  {author} {\bibinfo {author} {\bibfnamefont {A.~A.}\ \bibnamefont
  {Holmes}}, \bibinfo {author} {\bibfnamefont {N.~M.}\ \bibnamefont {Tubman}},\
  and\ \bibinfo {author} {\bibfnamefont {C.~J.}\ \bibnamefont {Umrigar}},\
  }\bibfield  {title} {\bibinfo {title} {Heat-bath configuration interaction:
  An efficient selected configuration interaction algorithm inspired by
  heat-bath sampling},\ }\href {https://doi.org/10.1021/acs.jctc.6b00407}
  {\bibfield  {journal} {\bibinfo  {journal} {J. Chem. Theory Comput.}\
  }\textbf {\bibinfo {volume} {12}},\ \bibinfo {pages} {3674} (\bibinfo {year}
  {2016})}\BibitemShut {NoStop}%
\bibitem [{\citenamefont {Baroni}\ and\ \citenamefont
  {Moroni}(1999)}]{baroni_reptation_1999}%
  \BibitemOpen
  \bibfield  {author} {\bibinfo {author} {\bibfnamefont {S.}~\bibnamefont
  {Baroni}}\ and\ \bibinfo {author} {\bibfnamefont {S.}~\bibnamefont
  {Moroni}},\ }\bibfield  {title} {\bibinfo {title} {Reptation {Quantum}
  {Monte} {Carlo}: {A} {Method} for {Unbiased} {Ground}-{State} {Averages} and
  {Imaginary}-{Time} {Correlations}},\ }\href
  {https://doi.org/10.1103/PhysRevLett.82.4745} {\bibfield  {journal} {\bibinfo
   {journal} {Phys. Rev. Lett.}\ }\textbf {\bibinfo {volume} {82}},\ \bibinfo
  {pages} {4745} (\bibinfo {year} {1999})}\BibitemShut {NoStop}%
\bibitem [{\citenamefont {Foulkes}\ \emph {et~al.}(2001)\citenamefont
  {Foulkes}, \citenamefont {Mitas}, \citenamefont {Needs},\ and\ \citenamefont
  {Rajagopal}}]{foulkes_quantum_2001}%
  \BibitemOpen
  \bibfield  {author} {\bibinfo {author} {\bibfnamefont {W.~M.~C.}\
  \bibnamefont {Foulkes}}, \bibinfo {author} {\bibfnamefont {L.}~\bibnamefont
  {Mitas}}, \bibinfo {author} {\bibfnamefont {R.~J.}\ \bibnamefont {Needs}},\
  and\ \bibinfo {author} {\bibfnamefont {G.}~\bibnamefont {Rajagopal}},\
  }\bibfield  {title} {\bibinfo {title} {Quantum {Monte} {Carlo} simulations of
  solids},\ }\href {https://doi.org/10.1103/RevModPhys.73.33} {\bibfield
  {journal} {\bibinfo  {journal} {Rev. Mod. Phys.}\ }\textbf {\bibinfo {volume}
  {73}},\ \bibinfo {pages} {33} (\bibinfo {year} {2001})}\BibitemShut {NoStop}%
\bibitem [{\citenamefont {Zhang}\ and\ \citenamefont
  {Krakauer}(2003)}]{zhang_quantum_2003}%
  \BibitemOpen
  \bibfield  {author} {\bibinfo {author} {\bibfnamefont {S.}~\bibnamefont
  {Zhang}}\ and\ \bibinfo {author} {\bibfnamefont {H.}~\bibnamefont
  {Krakauer}},\ }\bibfield  {title} {\bibinfo {title} {Quantum {Monte} {Carlo}
  {Method} using {Phase}-{Free} {Random} {Walks} with {Slater}
  {Determinants}},\ }\href {https://doi.org/10.1103/PhysRevLett.90.136401}
  {\bibfield  {journal} {\bibinfo  {journal} {Phys. Rev. Lett.}\ }\textbf
  {\bibinfo {volume} {90}},\ \bibinfo {pages} {136401} (\bibinfo {year}
  {2003})}\BibitemShut {NoStop}%
\bibitem [{\citenamefont {Booth}\ \emph {et~al.}(2009)\citenamefont {Booth},
  \citenamefont {Thom},\ and\ \citenamefont {Alavi}}]{booth_fermion_2009}%
  \BibitemOpen
  \bibfield  {author} {\bibinfo {author} {\bibfnamefont {G.~H.}\ \bibnamefont
  {Booth}}, \bibinfo {author} {\bibfnamefont {A.~J.~W.}\ \bibnamefont {Thom}},\
  and\ \bibinfo {author} {\bibfnamefont {A.}~\bibnamefont {Alavi}},\ }\bibfield
   {title} {\bibinfo {title} {Fermion {Monte} {Carlo} without fixed nodes: {A}
  game of life, death, and annihilation in {Slater} determinant space},\ }\href
  {https://doi.org/10.1063/1.3193710} {\bibfield  {journal} {\bibinfo
  {journal} {J. Chem. Phys.}\ }\textbf {\bibinfo {volume} {131}},\ \bibinfo
  {pages} {054106} (\bibinfo {year} {2009})}\BibitemShut {NoStop}%
\bibitem [{\citenamefont {Foyevtsova}\ \emph {et~al.}(2014)\citenamefont
  {Foyevtsova}, \citenamefont {Krogel}, \citenamefont {Kim}, \citenamefont
  {Kent}, \citenamefont {Dagotto},\ and\ \citenamefont
  {Reboredo}}]{foyevtsova_abinitio_2014}%
  \BibitemOpen
  \bibfield  {author} {\bibinfo {author} {\bibfnamefont {K.}~\bibnamefont
  {Foyevtsova}}, \bibinfo {author} {\bibfnamefont {J.~T.}\ \bibnamefont
  {Krogel}}, \bibinfo {author} {\bibfnamefont {J.}~\bibnamefont {Kim}},
  \bibinfo {author} {\bibfnamefont {P.~R.~C.}\ \bibnamefont {Kent}}, \bibinfo
  {author} {\bibfnamefont {E.}~\bibnamefont {Dagotto}},\ and\ \bibinfo {author}
  {\bibfnamefont {F.~A.}\ \bibnamefont {Reboredo}},\ }\bibfield  {title}
  {\bibinfo {title} {Ab initio {Quantum Monte Carlo Calculations of Spin
  Superexchange in Cuprates: The Benchmarking Case of}
  ${\mathrm{Ca}}_{2}{\mathrm{CuO}}_{3}$},\ }\href
  {https://doi.org/10.1103/PhysRevX.4.031003} {\bibfield  {journal} {\bibinfo
  {journal} {Phys. Rev. X}\ }\textbf {\bibinfo {volume} {4}},\ \bibinfo {pages}
  {031003} (\bibinfo {year} {2014})}\BibitemShut {NoStop}%
\bibitem [{\citenamefont {White}\ and\ \citenamefont
  {Martin}(1999)}]{white_ab_1999}%
  \BibitemOpen
  \bibfield  {author} {\bibinfo {author} {\bibfnamefont {S.~R.}\ \bibnamefont
  {White}}\ and\ \bibinfo {author} {\bibfnamefont {R.~L.}\ \bibnamefont
  {Martin}},\ }\bibfield  {title} {\bibinfo {title} {\textit{{Ab} initio}
  quantum chemistry using the density matrix renormalization group},\ }\href
  {https://doi.org/10.1063/1.478295} {\bibfield  {journal} {\bibinfo  {journal}
  {J. Chem. Phys.}\ }\textbf {\bibinfo {volume} {110}},\ \bibinfo {pages}
  {4127} (\bibinfo {year} {1999})}\BibitemShut {NoStop}%
\bibitem [{\citenamefont {Chan}\ and\ \citenamefont
  {Head-Gordon}(2002)}]{chan_highly_2002}%
  \BibitemOpen
  \bibfield  {author} {\bibinfo {author} {\bibfnamefont {G.~K.-L.}\
  \bibnamefont {Chan}}\ and\ \bibinfo {author} {\bibfnamefont {M.}~\bibnamefont
  {Head-Gordon}},\ }\bibfield  {title} {\bibinfo {title} {Highly correlated
  calculations with a polynomial cost algorithm: {A} study of the density
  matrix renormalization group},\ }\bibfield  {journal} {\bibinfo  {journal}
  {J. Chem. Phys.}\ }\textbf {\bibinfo {volume} {116}},\ \href
  {https://doi.org/10.1063/1.1449459} {10.1063/1.1449459} (\bibinfo {year}
  {2002})\BibitemShut {NoStop}%
\bibitem [{\citenamefont {Shepherd}\ \emph {et~al.}(2012)\citenamefont
  {Shepherd}, \citenamefont {Gr{\"u}neis}, \citenamefont {Booth}, \citenamefont
  {Kresse},\ and\ \citenamefont {Alavi}}]{shepherd_convergence_2012}%
  \BibitemOpen
  \bibfield  {author} {\bibinfo {author} {\bibfnamefont {J.~J.}\ \bibnamefont
  {Shepherd}}, \bibinfo {author} {\bibfnamefont {A.}~\bibnamefont
  {Gr{\"u}neis}}, \bibinfo {author} {\bibfnamefont {G.~H.}\ \bibnamefont
  {Booth}}, \bibinfo {author} {\bibfnamefont {G.}~\bibnamefont {Kresse}},\ and\
  \bibinfo {author} {\bibfnamefont {A.}~\bibnamefont {Alavi}},\ }\bibfield
  {title} {\bibinfo {title} {Convergence of many-body wave-function expansions
  using a plane-wave basis: {From} homogeneous electron gas to solid state
  systems},\ }\href {https://doi.org/10.1103/PhysRevB.86.035111} {\bibfield
  {journal} {\bibinfo  {journal} {Phys. Rev. B}\ }\textbf {\bibinfo {volume}
  {86}},\ \bibinfo {pages} {035111} (\bibinfo {year} {2012})}\BibitemShut
  {NoStop}%
\bibitem [{\citenamefont {Motta}\ \emph {et~al.}(2020)\citenamefont {Motta},
  \citenamefont {Genovese}, \citenamefont {Ma}, \citenamefont {Cui},
  \citenamefont {Sawaya}, \citenamefont {Chan}, \citenamefont {Chepiga},
  \citenamefont {Helms}, \citenamefont {Jim\'enez-Hoyos}, \citenamefont
  {Millis}, \citenamefont {Ray}, \citenamefont {Ronca}, \citenamefont {Shi},
  \citenamefont {Sorella}, \citenamefont {Stoudenmire}, \citenamefont {White},\
  and\ \citenamefont {Zhang}}]{motta_ground-state_2020}%
  \BibitemOpen
  \bibfield  {author} {\bibinfo {author} {\bibfnamefont {M.}~\bibnamefont
  {Motta}}, \bibinfo {author} {\bibfnamefont {C.}~\bibnamefont {Genovese}},
  \bibinfo {author} {\bibfnamefont {F.}~\bibnamefont {Ma}}, \bibinfo {author}
  {\bibfnamefont {Z.-H.}\ \bibnamefont {Cui}}, \bibinfo {author} {\bibfnamefont
  {R.}~\bibnamefont {Sawaya}}, \bibinfo {author} {\bibfnamefont {G.~K.-L.}\
  \bibnamefont {Chan}}, \bibinfo {author} {\bibfnamefont {N.}~\bibnamefont
  {Chepiga}}, \bibinfo {author} {\bibfnamefont {P.}~\bibnamefont {Helms}},
  \bibinfo {author} {\bibfnamefont {C.}~\bibnamefont {Jim\'enez-Hoyos}},
  \bibinfo {author} {\bibfnamefont {A.~J.}\ \bibnamefont {Millis}}, \bibinfo
  {author} {\bibfnamefont {U.}~\bibnamefont {Ray}}, \bibinfo {author}
  {\bibfnamefont {E.}~\bibnamefont {Ronca}}, \bibinfo {author} {\bibfnamefont
  {H.}~\bibnamefont {Shi}}, \bibinfo {author} {\bibfnamefont {S.}~\bibnamefont
  {Sorella}}, \bibinfo {author} {\bibfnamefont {E.~M.}\ \bibnamefont
  {Stoudenmire}}, \bibinfo {author} {\bibfnamefont {S.~R.}\ \bibnamefont
  {White}},\ and\ \bibinfo {author} {\bibfnamefont {S.}~\bibnamefont {Zhang}}
  (\bibinfo {collaboration} {Simons Collaboration on the Many-Electron
  Problem}),\ }\bibfield  {title} {\bibinfo {title} {Ground-{State}
  {Properties} of the {Hydrogen} {Chain}: {Dimerization},
  {Insulator}-to-{Metal} {Transition}, and {Magnetic} {Phases}},\ }\href
  {https://doi.org/10.1103/PhysRevX.10.031058} {\bibfield  {journal} {\bibinfo
  {journal} {Phys. Rev. X}\ }\textbf {\bibinfo {volume} {10}},\ \bibinfo
  {pages} {031058} (\bibinfo {year} {2020})}\BibitemShut {NoStop}%
\bibitem [{\citenamefont {Benali}\ \emph {et~al.}(2020)\citenamefont {Benali},
  \citenamefont {Gasperich}, \citenamefont {Jordan}, \citenamefont
  {Applencourt}, \citenamefont {Luo}, \citenamefont {Bennett}, \citenamefont
  {Krogel}, \citenamefont {Shulenburger}, \citenamefont {Kent}, \citenamefont
  {Loos}, \citenamefont {Scemama},\ and\ \citenamefont
  {Caffarel}}]{benali_toward_2020}%
  \BibitemOpen
  \bibfield  {author} {\bibinfo {author} {\bibfnamefont {A.}~\bibnamefont
  {Benali}}, \bibinfo {author} {\bibfnamefont {K.}~\bibnamefont {Gasperich}},
  \bibinfo {author} {\bibfnamefont {K.~D.}\ \bibnamefont {Jordan}}, \bibinfo
  {author} {\bibfnamefont {T.}~\bibnamefont {Applencourt}}, \bibinfo {author}
  {\bibfnamefont {Y.}~\bibnamefont {Luo}}, \bibinfo {author} {\bibfnamefont
  {M.~C.}\ \bibnamefont {Bennett}}, \bibinfo {author} {\bibfnamefont {J.~T.}\
  \bibnamefont {Krogel}}, \bibinfo {author} {\bibfnamefont {L.}~\bibnamefont
  {Shulenburger}}, \bibinfo {author} {\bibfnamefont {P.~R.~C.}\ \bibnamefont
  {Kent}}, \bibinfo {author} {\bibfnamefont {P.-F.}\ \bibnamefont {Loos}},
  \bibinfo {author} {\bibfnamefont {A.}~\bibnamefont {Scemama}},\ and\ \bibinfo
  {author} {\bibfnamefont {M.}~\bibnamefont {Caffarel}},\ }\bibfield  {title}
  {\bibinfo {title} {Toward a systematic improvement of the fixed-node
  approximation in diffusion {Monte} {Carlo} for solids—{A} case study in
  diamond},\ }\href {https://doi.org/10.1063/5.0021036} {\bibfield  {journal}
  {\bibinfo  {journal} {J. Chem. Phys.}\ }\textbf {\bibinfo {volume} {153}},\
  \bibinfo {pages} {184111} (\bibinfo {year} {2020})}\BibitemShut {NoStop}%
\bibitem [{\citenamefont {Szilva}\ \emph {et~al.}(2023)\citenamefont {Szilva},
  \citenamefont {Kvashnin}, \citenamefont {Stepanov}, \citenamefont
  {Nordstr{\"o}m}, \citenamefont {Eriksson}, \citenamefont {Lichtenstein},\
  and\ \citenamefont {Katsnelson}}]{szilva_quantitative_2023}%
  \BibitemOpen
  \bibfield  {author} {\bibinfo {author} {\bibfnamefont {A.}~\bibnamefont
  {Szilva}}, \bibinfo {author} {\bibfnamefont {Y.}~\bibnamefont {Kvashnin}},
  \bibinfo {author} {\bibfnamefont {E.~A.}\ \bibnamefont {Stepanov}}, \bibinfo
  {author} {\bibfnamefont {L.}~\bibnamefont {Nordstr{\"o}m}}, \bibinfo {author}
  {\bibfnamefont {O.}~\bibnamefont {Eriksson}}, \bibinfo {author}
  {\bibfnamefont {A.~I.}\ \bibnamefont {Lichtenstein}},\ and\ \bibinfo {author}
  {\bibfnamefont {M.~I.}\ \bibnamefont {Katsnelson}},\ }\bibfield  {title}
  {\bibinfo {title} {Quantitative theory of magnetic interactions in solids},\
  }\href {https://doi.org/10.1103/RevModPhys.95.035004} {\bibfield  {journal}
  {\bibinfo  {journal} {Rev. Mod. Phys.}\ }\textbf {\bibinfo {volume} {95}},\
  \bibinfo {pages} {035004} (\bibinfo {year} {2023})}\BibitemShut {NoStop}%
\bibitem [{Note1()}]{Note1}%
  \BibitemOpen
  \bibinfo {note} {For the simple case of \protect \textit {ab initio} hydrogen
  chain, spectral-fitting DMD also performs well in both cases, but could
  scramble the physics for more complicated system}\BibitemShut {NoStop}%
\bibitem [{\citenamefont {Dunning}(1989)}]{dunning_gaussian_1989}%
  \BibitemOpen
  \bibfield  {author} {\bibinfo {author} {\bibfnamefont {T.~H.}\ \bibnamefont
  {Dunning}},\ }\bibfield  {title} {\bibinfo {title} {Gaussian basis sets for
  use in correlated molecular calculations. {I}. {The} atoms boron through neon
  and hydrogen},\ }\href {https://doi.org/10.1063/1.456153} {\bibfield
  {journal} {\bibinfo  {journal} {J. Chem. Phys.}\ }\textbf {\bibinfo {volume}
  {90}},\ \bibinfo {pages} {1007} (\bibinfo {year} {1989})}\BibitemShut
  {NoStop}%
\bibitem [{\citenamefont {Wheeler}\ \emph {et~al.}(2023)\citenamefont
  {Wheeler}, \citenamefont {Pathak}, \citenamefont {Kleiner}, \citenamefont
  {Yuan}, \citenamefont {Rodrigues}, \citenamefont {Lorsung}, \citenamefont
  {Krongchon}, \citenamefont {Chang}, \citenamefont {Zhou}, \citenamefont
  {Busemeyer}, \citenamefont {Williams}, \citenamefont {Mu\~{n}oz},
  \citenamefont {Chow},\ and\ \citenamefont {Wagner}}]{wheeler_pyqmc_2022}%
  \BibitemOpen
  \bibfield  {author} {\bibinfo {author} {\bibfnamefont {W.~A.}\ \bibnamefont
  {Wheeler}}, \bibinfo {author} {\bibfnamefont {S.}~\bibnamefont {Pathak}},
  \bibinfo {author} {\bibfnamefont {K.~G.}\ \bibnamefont {Kleiner}}, \bibinfo
  {author} {\bibfnamefont {S.}~\bibnamefont {Yuan}}, \bibinfo {author}
  {\bibfnamefont {J.~N.~B.}\ \bibnamefont {Rodrigues}}, \bibinfo {author}
  {\bibfnamefont {C.}~\bibnamefont {Lorsung}}, \bibinfo {author} {\bibfnamefont
  {K.}~\bibnamefont {Krongchon}}, \bibinfo {author} {\bibfnamefont
  {Y.}~\bibnamefont {Chang}}, \bibinfo {author} {\bibfnamefont
  {Y.}~\bibnamefont {Zhou}}, \bibinfo {author} {\bibfnamefont {B.}~\bibnamefont
  {Busemeyer}}, \bibinfo {author} {\bibfnamefont {K.~T.}\ \bibnamefont
  {Williams}}, \bibinfo {author} {\bibfnamefont {A.}~\bibnamefont {Mu\~{n}oz}},
  \bibinfo {author} {\bibfnamefont {C.~Y.}\ \bibnamefont {Chow}},\ and\
  \bibinfo {author} {\bibfnamefont {L.~K.}\ \bibnamefont {Wagner}},\ }\bibfield
   {title} {\bibinfo {title} {{PyQMC: An all-Python real-space quantum Monte
  Carlo module in PySCF}},\ }\href {https://doi.org/10.1063/5.0139024}
  {\bibfield  {journal} {\bibinfo  {journal} {J. Chem. Phys.}\ }\textbf
  {\bibinfo {volume} {158}},\ \bibinfo {pages} {114801} (\bibinfo {year}
  {2023})}\BibitemShut {NoStop}%
\bibitem [{\citenamefont {Knizia}(2013)}]{knizia_intrinsic_2013}%
  \BibitemOpen
  \bibfield  {author} {\bibinfo {author} {\bibfnamefont {G.}~\bibnamefont
  {Knizia}},\ }\bibfield  {title} {\bibinfo {title} {Intrinsic {Atomic}
  {Orbitals}: {An} {Unbiased} {Bridge} between {Quantum} {Theory} and
  {Chemical} {Concepts}},\ }\href {https://doi.org/10.1021/ct400687b}
  {\bibfield  {journal} {\bibinfo  {journal} {J. Chem. Theory Comput.}\
  }\textbf {\bibinfo {volume} {9}},\ \bibinfo {pages} {4834} (\bibinfo {year}
  {2013})}\BibitemShut {NoStop}%
\bibitem [{\citenamefont {Reynolds}(2009)}]{reynolds_gaussian_2009}%
  \BibitemOpen
  \bibfield  {author} {\bibinfo {author} {\bibfnamefont {D.~A.}\ \bibnamefont
  {Reynolds}},\ }\href {https://doi.org/10.1007/978-1-4899-7488-4} {\emph
  {\bibinfo {title} {Gaussian {Mixture} {Models}}}},\ edited by\ \bibinfo
  {editor} {\bibfnamefont {S.~Z.}\ \bibnamefont {Li}}\ and\ \bibinfo {editor}
  {\bibfnamefont {A.~K.}\ \bibnamefont {Jain}},\ Vol.\ \bibinfo {volume} {741}\
  (\bibinfo  {publisher} {Boston, Springer},\ \bibinfo {year} {2009})\ pp.\
  \bibinfo {pages} {827--832}\BibitemShut {NoStop}%
\bibitem [{\citenamefont {Pedregosa}\ \emph {et~al.}(2011)\citenamefont
  {Pedregosa}, \citenamefont {Varoquaux}, \citenamefont {Gramfort},
  \citenamefont {Michel}, \citenamefont {Thirion}, \citenamefont {Grisel},
  \citenamefont {Blondel}, \citenamefont {Prettenhofer}, \citenamefont {Weiss},
  \citenamefont {Dubourg}, \citenamefont {Vanderplas}, \citenamefont {Passos},
  \citenamefont {Cournapeau}, \citenamefont {Brucher}, \citenamefont {Perrot},\
  and\ \citenamefont {Duchesnay}}]{pedregosa_scikit-learn_2011}%
  \BibitemOpen
  \bibfield  {author} {\bibinfo {author} {\bibfnamefont {F.}~\bibnamefont
  {Pedregosa}}, \bibinfo {author} {\bibfnamefont {G.}~\bibnamefont
  {Varoquaux}}, \bibinfo {author} {\bibfnamefont {A.}~\bibnamefont {Gramfort}},
  \bibinfo {author} {\bibfnamefont {V.}~\bibnamefont {Michel}}, \bibinfo
  {author} {\bibfnamefont {B.}~\bibnamefont {Thirion}}, \bibinfo {author}
  {\bibfnamefont {O.}~\bibnamefont {Grisel}}, \bibinfo {author} {\bibfnamefont
  {M.}~\bibnamefont {Blondel}}, \bibinfo {author} {\bibfnamefont
  {P.}~\bibnamefont {Prettenhofer}}, \bibinfo {author} {\bibfnamefont
  {R.}~\bibnamefont {Weiss}}, \bibinfo {author} {\bibfnamefont
  {V.}~\bibnamefont {Dubourg}}, \bibinfo {author} {\bibfnamefont
  {J.}~\bibnamefont {Vanderplas}}, \bibinfo {author} {\bibfnamefont
  {A.}~\bibnamefont {Passos}}, \bibinfo {author} {\bibfnamefont
  {D.}~\bibnamefont {Cournapeau}}, \bibinfo {author} {\bibfnamefont
  {M.}~\bibnamefont {Brucher}}, \bibinfo {author} {\bibfnamefont
  {M.}~\bibnamefont {Perrot}},\ and\ \bibinfo {author} {\bibfnamefont
  {E.}~\bibnamefont {Duchesnay}},\ }\bibfield  {title} {\bibinfo {title}
  {Scikit-learn: {Machine} {Learning} in {P}ython},\ }\href
  {https://www.jmlr.org/papers/volume12/pedregosa11a/pedregosa11a.pdf?ref=https:/}
  {\bibfield  {journal} {\bibinfo  {journal} {J. Mach. Learn. Res.}\ }\textbf
  {\bibinfo {volume} {12}},\ \bibinfo {pages} {2825} (\bibinfo {year}
  {2011})}\BibitemShut {NoStop}%
\bibitem [{\citenamefont {Schwarz}(1978)}]{schwarz_estimating_1978}%
  \BibitemOpen
  \bibfield  {author} {\bibinfo {author} {\bibfnamefont {G.}~\bibnamefont
  {Schwarz}},\ }\bibfield  {title} {\bibinfo {title} {Estimating the
  {Dimension} of a {Model}},\ }\bibfield  {journal} {\bibinfo  {journal} {Ann.
  Statist.}\ }\textbf {\bibinfo {volume} {6}},\ \href
  {https://doi.org/10.1214/aos/1176344136} {10.1214/aos/1176344136} (\bibinfo
  {year} {1978})\BibitemShut {NoStop}%
\bibitem [{\citenamefont {Kodinariya}\ and\ \citenamefont
  {Makwana}(2013)}]{kodinariya_review_2013}%
  \BibitemOpen
  \bibfield  {author} {\bibinfo {author} {\bibfnamefont {T.~M.}\ \bibnamefont
  {Kodinariya}}\ and\ \bibinfo {author} {\bibfnamefont {P.~R.}\ \bibnamefont
  {Makwana}},\ }\bibfield  {title} {\bibinfo {title} {Review on determining
  number of cluster in k-means clustering},\ }\href
  {https://www.researchgate.net/profile/Trupti-Kodinariya/publication/313554124_Review_on_Determining_of_Cluster_in_K-means_Clustering/links/5789fda408ae59aa667931d2/Review-on-Determining-of-Cluster-in-K-means-Clustering.pdf}
  {\bibfield  {journal} {\bibinfo  {journal} {Int. J.}\ }\textbf {\bibinfo
  {volume} {1}},\ \bibinfo {pages} {90} (\bibinfo {year} {2013})}\BibitemShut
  {NoStop}%
\bibitem [{\citenamefont {Kuhn}(1955)}]{kuhn_hungarian_1955}%
  \BibitemOpen
  \bibfield  {author} {\bibinfo {author} {\bibfnamefont {H.~W.}\ \bibnamefont
  {Kuhn}},\ }\bibfield  {title} {\bibinfo {title} {The {Hungarian} method for
  the assignment problem},\ }\href {https://doi.org/10.1002/nav.3800020109}
  {\bibfield  {journal} {\bibinfo  {journal} {Nav. Res. Logist. Q.}\ }\textbf
  {\bibinfo {volume} {2}},\ \bibinfo {pages} {83} (\bibinfo {year}
  {1955})}\BibitemShut {NoStop}%
\bibitem [{\citenamefont {Shoji}\ \emph {et~al.}(2006)\citenamefont {Shoji},
  \citenamefont {Koizumi}, \citenamefont {Kitagawa}, \citenamefont {Kawakami},
  \citenamefont {Yamanaka}, \citenamefont {Okumura},\ and\ \citenamefont
  {Yamaguchi}}]{shoji_general_2006}%
  \BibitemOpen
  \bibfield  {author} {\bibinfo {author} {\bibfnamefont {M.}~\bibnamefont
  {Shoji}}, \bibinfo {author} {\bibfnamefont {K.}~\bibnamefont {Koizumi}},
  \bibinfo {author} {\bibfnamefont {Y.}~\bibnamefont {Kitagawa}}, \bibinfo
  {author} {\bibfnamefont {T.}~\bibnamefont {Kawakami}}, \bibinfo {author}
  {\bibfnamefont {S.}~\bibnamefont {Yamanaka}}, \bibinfo {author}
  {\bibfnamefont {M.}~\bibnamefont {Okumura}},\ and\ \bibinfo {author}
  {\bibfnamefont {K.}~\bibnamefont {Yamaguchi}},\ }\bibfield  {title} {\bibinfo
  {title} {A general algorithm for calculation of {Heisenberg} exchange
  integrals {$J$} in multispin systems},\ }\href
  {https://doi.org/10.1016/j.cplett.2006.10.023} {\bibfield  {journal}
  {\bibinfo  {journal} {Chem. Phys. Lett.}\ }\textbf {\bibinfo {volume}
  {432}},\ \bibinfo {pages} {343} (\bibinfo {year} {2006})}\BibitemShut
  {NoStop}%
\bibitem [{\citenamefont {Schrieffer}\ and\ \citenamefont
  {Wolff}(1966)}]{schrieffer_relation_1966}%
  \BibitemOpen
  \bibfield  {author} {\bibinfo {author} {\bibfnamefont {J.~R.}\ \bibnamefont
  {Schrieffer}}\ and\ \bibinfo {author} {\bibfnamefont {P.~A.}\ \bibnamefont
  {Wolff}},\ }\bibfield  {title} {\bibinfo {title} {Relation between the
  {Anderson} and {Kondo} {Hamiltonians}},\ }\href
  {https://doi.org/10.1103/PhysRev.149.491} {\bibfield  {journal} {\bibinfo
  {journal} {Phys. Rev.}\ }\textbf {\bibinfo {volume} {149}},\ \bibinfo {pages}
  {491} (\bibinfo {year} {1966})}\BibitemShut {NoStop}%
\bibitem [{\citenamefont {Bravyi}\ \emph {et~al.}(2011)\citenamefont {Bravyi},
  \citenamefont {DiVincenzo},\ and\ \citenamefont
  {Loss}}]{bravyi_schriefferwolff_2011}%
  \BibitemOpen
  \bibfield  {author} {\bibinfo {author} {\bibfnamefont {S.}~\bibnamefont
  {Bravyi}}, \bibinfo {author} {\bibfnamefont {D.~P.}\ \bibnamefont
  {DiVincenzo}},\ and\ \bibinfo {author} {\bibfnamefont {D.}~\bibnamefont
  {Loss}},\ }\bibfield  {title} {\bibinfo {title} {Schrieffer–{Wolff}
  transformation for quantum many-body systems},\ }\href
  {https://doi.org/10.1016/j.aop.2011.06.004} {\bibfield  {journal} {\bibinfo
  {journal} {Ann. Phys.}\ }\textbf {\bibinfo {volume} {326}},\ \bibinfo {pages}
  {2793} (\bibinfo {year} {2011})}\BibitemShut {NoStop}%
\bibitem [{\citenamefont {Kuzmenko}\ \emph {et~al.}(2009)\citenamefont
  {Kuzmenko}, \citenamefont {Crassee}, \citenamefont {van~der Marel},
  \citenamefont {Blake},\ and\ \citenamefont
  {Novoselov}}]{kuzmenko_determination_2009}%
  \BibitemOpen
  \bibfield  {author} {\bibinfo {author} {\bibfnamefont {A.~B.}\ \bibnamefont
  {Kuzmenko}}, \bibinfo {author} {\bibfnamefont {I.}~\bibnamefont {Crassee}},
  \bibinfo {author} {\bibfnamefont {D.}~\bibnamefont {van~der Marel}}, \bibinfo
  {author} {\bibfnamefont {P.}~\bibnamefont {Blake}},\ and\ \bibinfo {author}
  {\bibfnamefont {K.~S.}\ \bibnamefont {Novoselov}},\ }\bibfield  {title}
  {\bibinfo {title} {Determination of the gate-tunable band gap and
  tight-binding parameters in bilayer graphene using infrared spectroscopy},\
  }\href {https://doi.org/10.1103/PhysRevB.80.165406} {\bibfield  {journal}
  {\bibinfo  {journal} {Phys. Rev. B}\ }\textbf {\bibinfo {volume} {80}},\
  \bibinfo {pages} {165406} (\bibinfo {year} {2009})}\BibitemShut {NoStop}%
\bibitem [{\citenamefont {Ribeiro}\ \emph {et~al.}(2009)\citenamefont
  {Ribeiro}, \citenamefont {Pereira}, \citenamefont {Peres}, \citenamefont
  {Briddon},\ and\ \citenamefont {Castro~Neto}}]{ribeiro_strained_2009}%
  \BibitemOpen
  \bibfield  {author} {\bibinfo {author} {\bibfnamefont {R.~M.}\ \bibnamefont
  {Ribeiro}}, \bibinfo {author} {\bibfnamefont {V.~M.}\ \bibnamefont
  {Pereira}}, \bibinfo {author} {\bibfnamefont {N.~M.~R.}\ \bibnamefont
  {Peres}}, \bibinfo {author} {\bibfnamefont {P.~R.}\ \bibnamefont {Briddon}},\
  and\ \bibinfo {author} {\bibfnamefont {A.~H.}\ \bibnamefont {Castro~Neto}},\
  }\bibfield  {title} {\bibinfo {title} {Strained graphene: tight-binding and
  density functional calculations},\ }\href
  {https://doi.org/10.1088/1367-2630/11/11/115002} {\bibfield  {journal}
  {\bibinfo  {journal} {New J. Phys.}\ }\textbf {\bibinfo {volume} {11}},\
  \bibinfo {pages} {115002} (\bibinfo {year} {2009})}\BibitemShut {NoStop}%
\bibitem [{\citenamefont {Changlani}\ \emph {et~al.}(2015)\citenamefont
  {Changlani}, \citenamefont {Zheng},\ and\ \citenamefont
  {Wagner}}]{changlani_density-matrix_2015}%
  \BibitemOpen
  \bibfield  {author} {\bibinfo {author} {\bibfnamefont {H.~J.}\ \bibnamefont
  {Changlani}}, \bibinfo {author} {\bibfnamefont {H.}~\bibnamefont {Zheng}},\
  and\ \bibinfo {author} {\bibfnamefont {L.~K.}\ \bibnamefont {Wagner}},\
  }\bibfield  {title} {\bibinfo {title} {Density-matrix based determination of
  low-energy model {Hamiltonians} from {\textit{ab initio}} wavefunctions},\
  }\href {https://doi.org/10.1063/1.4927664} {\bibfield  {journal} {\bibinfo
  {journal} {J. Chem. Phys.}\ }\textbf {\bibinfo {volume} {143}},\ \bibinfo
  {pages} {102814} (\bibinfo {year} {2015})}\BibitemShut {NoStop}%
\bibitem [{\citenamefont {Zheng}\ \emph {et~al.}(2018)\citenamefont {Zheng},
  \citenamefont {Changlani}, \citenamefont {Williams}, \citenamefont
  {Busemeyer},\ and\ \citenamefont {Wagner}}]{zheng_real_2018}%
  \BibitemOpen
  \bibfield  {author} {\bibinfo {author} {\bibfnamefont {H.}~\bibnamefont
  {Zheng}}, \bibinfo {author} {\bibfnamefont {H.~J.}\ \bibnamefont
  {Changlani}}, \bibinfo {author} {\bibfnamefont {K.~T.}\ \bibnamefont
  {Williams}}, \bibinfo {author} {\bibfnamefont {B.}~\bibnamefont
  {Busemeyer}},\ and\ \bibinfo {author} {\bibfnamefont {L.~K.}\ \bibnamefont
  {Wagner}},\ }\bibfield  {title} {\bibinfo {title} {From {Real} {Materials} to
  {Model} {Hamiltonians} {With} {Density} {Matrix} {Downfolding}},\ }\href
  {https://doi.org/10.3389/fphy.2018.00043} {\bibfield  {journal} {\bibinfo
  {journal} {Front. Phys.}\ }\textbf {\bibinfo {volume} {6}},\ \bibinfo {pages}
  {43} (\bibinfo {year} {2018})}\BibitemShut {NoStop}%
\bibitem [{\citenamefont {Chang}\ and\ \citenamefont
  {Wagner}(2020)}]{chang_effective_2020}%
  \BibitemOpen
  \bibfield  {author} {\bibinfo {author} {\bibfnamefont {Y.}~\bibnamefont
  {Chang}}\ and\ \bibinfo {author} {\bibfnamefont {L.~K.}\ \bibnamefont
  {Wagner}},\ }\bibfield  {title} {\bibinfo {title} {Effective spin-orbit
  models using correlated first-principles wave functions},\ }\href
  {https://doi.org/10.1103/PhysRevResearch.2.013195} {\bibfield  {journal}
  {\bibinfo  {journal} {Phys. Rev. Research}\ }\textbf {\bibinfo {volume}
  {2}},\ \bibinfo {pages} {013195} (\bibinfo {year} {2020})}\BibitemShut
  {NoStop}%
\bibitem [{\citenamefont {Sawaya}\ and\ \citenamefont
  {White}(2022)}]{sawaya_constructing_2022}%
  \BibitemOpen
  \bibfield  {author} {\bibinfo {author} {\bibfnamefont {R.~C.}\ \bibnamefont
  {Sawaya}}\ and\ \bibinfo {author} {\bibfnamefont {S.~R.}\ \bibnamefont
  {White}},\ }\bibfield  {title} {\bibinfo {title} {Constructing {Hubbard}
  models for the hydrogen chain using sliced-basis density matrix
  renormalization group},\ }\href {https://doi.org/10.1103/PhysRevB.105.045145}
  {\bibfield  {journal} {\bibinfo  {journal} {Phys. Rev. B}\ }\textbf {\bibinfo
  {volume} {105}},\ \bibinfo {pages} {045145} (\bibinfo {year}
  {2022})}\BibitemShut {NoStop}%
\bibitem [{\citenamefont {Mallat}\ and\ \citenamefont {{Zhifeng
  Zhang}}(1993)}]{mallat_matching_1993}%
  \BibitemOpen
  \bibfield  {author} {\bibinfo {author} {\bibfnamefont {S.}~\bibnamefont
  {Mallat}}\ and\ \bibinfo {author} {\bibnamefont {{Zhifeng Zhang}}},\
  }\bibfield  {title} {\bibinfo {title} {Matching pursuits with time-frequency
  dictionaries},\ }\href {https://doi.org/10.1109/78.258082} {\bibfield
  {journal} {\bibinfo  {journal} {IEEE Trans. Signal Process.}\ }\textbf
  {\bibinfo {volume} {41}},\ \bibinfo {pages} {3397} (\bibinfo {year}
  {1993})}\BibitemShut {NoStop}%
\end{thebibliography}
\end{document}